\documentclass[useAMS,usenatbib,usegraphicx,twocolumn]{mn2e}
  
\usepackage{float}
\usepackage{aas_macros}
\usepackage{threeparttable}
\usepackage{grffile}
\usepackage{amsmath, amssymb}
\usepackage{multirow}
\usepackage{wallpaper}
\usepackage{mathbbol}
\usepackage{rotating}
\usepackage{blindtext}
\usepackage[T1]{fontenc}
\usepackage{changes}
\usepackage{lscape}

\usepackage{tocloft}
\newcommand{\fat}{\textbf}
\newcommand{\ita}{\textit}

\newcommand{\beq}{\begin{equation}}
\newcommand{\eeq}{\end{equation}}  
\newcommand{\RNum}[1]{\uppercase\expandafter{\romannumeral #1\relax}}

\newcommand{\Hone}{\rm{H}\textsc{i} }
\newcommand{\bk}[1]{{\bf\textcolor{red}{#1}}}
\newcommand{\np}[1]{{\bf\textcolor{red}{#1}}}
\newcommand{\bkk}[1]{{\bf\textcolor{red}{#1}}}
\renewcommand{\bk}[1]{#1}
\renewcommand{\bkk}[1]{#1}
\renewcommand{\np}[1]{#1}
  \title[]{The Spatial Power Spectrum and Derived Turbulent Properties of 
  Isolated Galaxies}
  
  \author[K\"ortgen, Pingel \& Killerby-Smith]{Bastian~K\"ortgen$^1$, Nickolas Pingel$^2$ and Nicholas Killerby-Smith$^2$\\
  $^1$ Hamburger Sternwarte, Universit\"at Hamburg, Gojenbergsweg 112, D-21029 Hamburg, Germany \\
  $^2$ Research School of Astronomy \& Astrophysics, Australian National University, Canberra, ACT 2611, Australia \\
  }

\date{Released 2021}

\pagerange{\pageref{firstpage}--\pageref{lastpage}} \pubyear{2021}

\begin{document}

\label{firstpage}
\maketitle

\begin{abstract}
The turbulent dynamics of nearby and extragalactic gas structures 
can be studied with the \bk{column} density power spectrum, which is
often described by a broken power-law. In an extragalactic context, the breaks in 
the power spectra have been interpreted to constrain the disc scale height\bk{, which marks a transition from 2D disc-like to 3D motion}. However,  this 
interpretation has recently been questioned when accounting for instrumental effects. We use numerical simulations to study the spatial power spectra of isolated galaxies and investigate the origins of the break scale. We split the gas into various phases and analyze the time evolution of the power spectrum characteristics, such as the slope(s) and the break scale. We find that the break scale is phase dependent. The physics traced by the break scale also differ: in the warm gas it 
marks the transition from 2D (disk--like) to 3D (isotropic) turbulence. In the cold gas, the break scale traces the typical size of molecular clouds. We further show that the break scale almost never traces the disc scale height. We study turbulent properties of the ISM to show that, in the case where the break scale traces a transition to isotropic turbulence, the fraction of required accretion energy to sustain turbulent motions in the ISM increases significantly. Lastly, we demonstrate through simulated observations that it is crucial to account for observational effects, such as the beam and instrumental noise, in order to accurately recover the break scale in real observations. 
\end{abstract}

\section{Introduction}
Turbulence is ubiquitous in the interstellar medium (ISM) 
of galaxies (e.g., \citealt{Elmegreen04}). It plays an essential role for various 
physical processes, such as magnetic field 
amplification \citep{Federrath2016_plasma}, the mixing of metals \citep{Yang2012}, or the formation of stars 
within gravitationally contracting molecular clouds \citep{Burkhart2015}. Hence, 
understanding the properties and statistics of interstellar 
turbulence is of utmost importance for understanding the 
above mentioned processes and the overall dynamics of the 
ISM.\\
Theoretically, turbulence can be characterized by various statistical metrics, 
such as e.g. the properties of the density probability distribution 
function (PDF; \citealt{Federrath08, burkhart2017}) or power spectra and structure functions of density (e.g., \citealt{stanimirovic1999, Muller2004, pingel2013, pingel2018}), and  
velocity or kinetic energy \citep{Grisdale2017}. Since turbulence in the ISM is generally 
trans-sonic and compressible \citep[e.g.,][]{Elmegreen04}, the latter metric can be applied to 
both compressive and solenoidal velocity modes after decomposing 
the velocity field.\\
Observationally, the properties of interstellar turbulence are impossible to assess in full detail. Thus, statistical 
approaches are chosen to compare observational results with 
predictions from (full 3D) theoretical or numerical models \citep[e.g.,][]{Scalo04,MacLow04,Federrath12}.
One typically assumes isotropy of the dynamics to infer statistics of e.g. the turbulent 
Mach number from the line-of-sight (LOS) velocity component or the (3D) volume density 
PDF from the (2D) column density PDF \citep{Brunt10c,Brunt10b}. Additionally,  
the index of the power spectrum of column density or intensity can be used to derive properties 
of the turbulent velocity field. This is valid under the 
assumption that the density field can be treated as a passively advected scalar, 
where the velocity and density spectra will have comparable shapes \citep[e.g.][]{Stanimirovic2001}. Using this approach, \citet{Stanimirovic2001} showed that 
the slope of \Hone intensity power spectra matches compressible, Burgers type, turbulence. \citet{pingel2018} used the column density power spectrum (or spatial power spectrum, SPS) and found 
that turbulence near the Perseus molecular cloud is characterized by almost 
trans-sonic motions in \Hone and mildly supersonic velocities in CO. The shallower observed slopes for CO and dust have been interpreted as a 
signature of gravitational collapse \citep{Burkhart2015}. \citet{Marchal2021} found that the WNM, observed as part of the GHIGLS survey \citep{Martin2015}, is trans-sonic and shows a LOS velocity and 
spatial power spectrum with slope $\sim-11/3$, i.e. incompressible Kolmogorov 
turbulence. \citet[][\bk{see also \citet{Nestingen17}}]{Szotkowski2019} studied \Hone SPS in the 
Magellanic clouds and found that SPS properties, such as 
the slopes on large and small scales do not vary much across the 
galaxies. In case of the LMC, they found some variation near 
30\,Doradus, due to intense stellar feedback, and towards the 
outskirts of the galaxy, which the authors interpreted as disc 
flaring.\\
Apart from the properties of the turbulence, galactic disc parameters may be derived from the SPS as well. If the SPS contains a break, i.e. a change in its slope, it is indicative of a characteristic scale within the observed system. \citet{Elmegreen2001}, \citet{Dutta2008} and \citet{Dutta2009} studied the SPS in several galaxies using various tracers, including optical and radio bands, and found a break at scales of $\sim100\,\mathrm{pc}$. As  \citet{Elmegreen2001} state, this is of the order of the disc scale height \bk{\citep[see similar arguments by][who use the Spectral Correlation Function]{Padoan2001}}. This has led to the interpretation that a break in the SPS traces the third dimension / the extent of face-on galaxies along the LOS. \citet{Bournaud2010} studied numerical simulations of isolated galaxies and found that the break in their simulated SPS is located at scales that correspond to the disc Jeans length. As the authors further argued, the Jeans length of the system is comparable to the disc scale height. More recent numerical investigations by \citet{Grisdale2017} have shown that the SPS is quite sensitive to gas layers between the galaxy and the observer, and that the column density of this layer can hide the 
presence of a break. In addition, \citet{koch2020} studied the SPS of various species in several galaxies. The authors demonstrated that the breaks previously measured in dust surface density SPS profiles of their sample can be fully characterized by foward-modelling of the instrumental point spread function \bk{\citep[PSF. See also][who compare their data to \Hone data from the THINGS survey.]{Grisdale2017}}. They concluded that there are multiple possible sources of influence on the presence of a break in SPS, including: the large-scale distribution of the gas, the presence of dense H$_{2}$-dominated regions, the gas tracer used, and instrumental systematics. \\
In this study, we utilize a simulation of the different gas phases of an isolated, face-on galaxy to investigate which properties of galaxies influence the shape of the SPS and existence of a break. The manuscript is ordered as follows: In Sec.~\ref{sec:methods} 
we explain the numerical code and initial conditions as well as 
how we compute the SPS; in Sec.~\ref{sec:results} we present and 
discuss our findings; Sec.~\ref{sec:discussion} discusses the characteristic length scales traced by the SPS and demonstrates the recovery of a break scale in simulated observations; this study is then closed with a summary in Sec.~\ref{sec:summary}.


\section{Methodology}\label{sec:methods}
\subsection{Numerical Method}
For our simulations of galaxy evolution we use the finite volume code \textsc{flash} \citep[v4.2.2,][]{Dubey08}. During each timestep, the equations of ideal magnetohydrodynamics are solved with a five-wave Riemann solver \citep{Bouchut09,Waagan11} on an adaptive mesh \citep[the AMR technique,][]{Berger84}. In addition, Poisson's equation for the self-gravity of 
the gas is solved with a tree-solver \citep[see e.g.][]{Lukat16}. We use optically thin heating and cooling rates, 
where the former is kept constant and the latter are provided in tabulated form and based upon the fitting formulae by 
\citet[][with modifications by \cite{Vazquez07}]{Koyama02}.\\
In order to follow gravitational contraction of cool, overdense gas, the 
numerical grid is adaptively refined once the local Jeans length is resolved 
with less than 32 grid cells and de-refined when it is made up by more than 64 
cells. The root grid is set to a resolution of \mbox{$\Delta x_\mathrm{root}=625\,\mathrm{pc}$} and the subsequent refinement steps yield a 
maximum resolution of \mbox{$\Delta x=19.5\,\mathrm{pc}$}. To ensure that no 
artificial fragmentation of the gas occurs, we incorporate an artificial 
pressure term on the highest level of refinement. Hence, the local Jeans length 
is still resolved by at least four grid cells \citep{Truelove97}. As provided in 
the appendix of \citet{Koertgen19b}, this set of resolution criteria implies that the disc scale height is well resolved from $R\sim5\,\mathrm{kpc}$ on and marginally resolved below this distance to the center. We do not include stellar feedback in this 
study to focus solely on the impact of the galaxy dynamics.

\subsection{Initial Conditions}
The disc is set up with a radially and vertically declining density 
profile of the form \citep{Tasker09,Koertgen18L}:
\beq
\varrho\left(R,z\right)=\frac{\kappa c_\mathrm{s}\sqrt{1+\frac{2}{\beta}}}{\pi G Q H(R)}\mathrm{sech}\left(\frac{z}{H(R)}\right).
\eeq
Here, $R$ and $z$ are the radial and vertical coordinates, $\kappa$ is 
the epicyclic frequency, $c_\mathrm{s}$ the speed of sound, $\beta$ and 
$Q$ denote the ratio of thermal to magnetic pressure and the Toomre 
stability parameter, and $H(R)$ is the radially varying disc 
scale height. The parameters were chosen in such way that $Q=2$ in the 
main part of the disc ($0.5<R/\mathrm{kpc}<8$) and $Q=20$ elsewhere\footnote{This high stability parameter is primarily for numerical reasons.}. The external logarithmic potential 
\beq
\Phi_\mathrm{ext}=\frac{1}{2}v_0^2\mathrm{ln}\left\{\frac{1}{R_\mathrm{c}^2}\left[R^2+R_\mathrm{c}^2+\left(\frac{z}{q}\right)^2\right]\right\}
\eeq
provides a flat rotation 
curve with an overall rotation speed of $v_0=200\,\mathrm{km/s}$ \citep[see also][]{Dobbs06}. We emphasize that, for reasons of simplicity, we do not add initial turbulent velocity fluctuations on top of the rotation velocity.\\ Galaxies are magnetized \citep{Crutcher12} and, hence, we add an initially toroidal magnetic field, which varies in strength as a function of density. This dependence ensures that the plasma-$\beta$ is constant 
throughout the main disc. For the galaxy studied below, we use an initial value of $\beta=0.25$. 

\subsection{Separation \bk{i}nto Phases}
Emission from \Hone gas is usually composed of contributions from the warm ($10000\,\mathrm{K}\,>\,T\,>\,5000\,\mathrm{K}$) and cold phases ($300\,\mathrm{K}\,>\,T\,>\,50\,\mathrm{K}$), which are frequency decomposed into separate contributions in observations via a Gaussian decomposition \citep[e.g.][]{Kalberla2018,Marchal2021}. For this reason, we separate the gas into distinct phases, based on their 
temperatures. Our prescription for heating and cooling \citep{Koyama02,Vazquez07} assigns a specific 
density to each given temperature so that a second criterion based on density is not necessary. Our choices 
for the various phases are given in Tab.~\ref{tab:phases}. In addition to the typical cold (CNM) and warm neutral phases (WNM), we define a transitional (TRA) phase that captures gas between the typical temperature ranges, and a dense H$_2$-dominated phase that has been observed to influence the small-scale components of measured SPS profiles. \bk{Apart from the individual phases, we further show their combination, which we term \ita{NEU} and \ita{TOT}, where the former encompasses all neutral gas (CNM+TRA+WNM) phases except for the coldest and densest phase, MOL, and the latter reflects the total column density of material including diffuse and dense gas.} \np{The addition of the NEU phase provides an important comparison to real observations of \Hone gas in nearby galaxies, such as the Hydrogen Accretion in LOcal GAlaxieS (HALOGAS) Survey with the Westerbork Synthesis Radio Telescope \citep{heald11}. Furthermore, the uniform mixing of the total \Hone column with dust \citep{galliano2018} enables an approximate extension of our results to another key component of the ISM.}
\begin{table*}
    \centering
    \caption{Definition of the various phases used 
    in this study. Temperature ranges are chosen 
    to be in broad agreement with the ranges defined by, e.g.,
    \citet{Wolfire95}, \citet{Wolfire03}, \citet{Vazquez12}, \citet{Kalberla2018} and \citet{Bracco20}.}
    \begin{tabular}{lccl}
        \hline
        \hline
        \fat{Phase key} &\fat{Upper temperature limit} &\fat{Lower temperature limit} &\fat{Comment}\\
        &$\left[\mathrm{K}\right]$  &$\left[\mathrm{K}\right]$  &\\
        \hline
         \fat{WNM}&8000&1000&Warm \Hone / Warm neutral medium (WNM)  \\
         \fat{TRA}&1000&300&Transitional gas / Lukewarm neutral medium (LNM) \\
         \fat{CNM}&300&50&Cold \Hone / Cold neutral medium (CNM) \\
         \fat{\bk{MOL}}&50&2&H$_2$ / molecular gas. 2\,K is our assigned temperature floor\\
          \bk{\fat{NEU}}&\bk{8000}&\bk{50}&\bk{Total \Hone gas (CNM+TRA+WNM)}\\
     \bk{\fat{TOT}}&\bk{8000}&\bk{2}&\np{Total gas content (MOL+CNM+TRA+WNM)} \\
         \hline
         \hline
    \end{tabular}
    \label{tab:phases}
\end{table*}
\subsection{Computing the Spatial Power Spectra}
The spatial power spectrum (SPS) is a widely used statistical diagnostic in the investigation into the turbulent properties of the ISM \citep[e.g.][and references therein]{Stanimirovic2001,Dutta2008,pingel2018,koch2020}. The SPS is formally the Fourier transform of the two-point auto-correlation function defined as
\begin{equation}
    P\left(k\right) = \mathfrak{F}\left(I\right)\times\mathfrak{F}^{*}\left(I\right), 
\end{equation}
where $\mathfrak{F}\left(I\right)$ is the Fourier transform of a simulated or observed 2D intensity field, $I$. In practice, we utilize the {\tt PowerSpectrum} object within the {\tt Turbustat}\footnote{\url{https://github.com/Astroua/TurbuStat/releases}} python package \citep{koch2019_turbustat} to construct a 1D radial SPS profile. We relate the spatial frequencies, $k$, to physical scales by
\begin{equation}
    k [\lambda] = \frac{1}{d [\mathrm{pc}]}, 
\end{equation}
where $d$ is the physical scale of the simulation. To ensure a sufficient number of pixels in the smallest spatial frequency bin, we probe physical scales limited to 1/2 the image size ($d_{\rm max}$ = 10240 pc; $k_{\rm min}$ = 9.77$\times$10$^{-5}\lambda$) down to adjacent pixels ($d_{\rm min}$ = 40 pc; $k_{max}$ = 0.025$\lambda$).\\
Most applications of the SPS compute the mean power value within each spatial frequency bin. However, taking the Fourier transform of an image that contains intensity all the way to its edges will exhibit the Gibb's phenomenon, where $sinc$ ringing occurs along the axes of the Fourier transform image. These outlying power values will subsequently bias the mean, especially at bins of small spatial frequencies. Several studies have reduced this ringing by applying apodization kernels that bring the intensity smoothly to zero at the edges. However, this approach risks losing information at specific scales. \np{We note, however, that our simulated images contain the full disk such that no low-level emission exists at the edge which may affect our analysis. In order to ensure information at all scales is preserved, we repeat our analysis taking the median power value within each spatial frequency bin. This approach in general will avoid outlying power values and better recover the true location of the peak of each distribution of power values for a given spatial frequency bin. See Fig. 4 in Section 3.1 of \citet{pingel2018} for a statistical demonstration. Repeating the analysis utilizing the median does not alter the resulting fits in any statistically meaningful way. Furthermore, the 2D SPS images do not show strong pixels along the central, which is the signature manifestation of Gibb's ringing, indicating our results are not biased by this effect. We therefore bin by the mean power value in each radial profile, as this is the default behaviour of {\tt Turbustat}.} 

The 1D profile is then fit in log-space via an ordinary least-squares algorithm to determine the power spectral slope. The uncertainties of the fit parameters are determined using a bootstrap method, which is the default scheme implemented in {\tt Turbustat}. Explicitly, the residuals from the fit are re-sampled and added back into the data. The re-sampled data are then re-fit for a total of 100 iterations to build up statistical distributions of the fit parameters. The final uncertainties for each fit parameter are then the standard deviation of the respective distribution. 

\subsection{Estimating the Break Scale}
In general, the power spectra of the ISM observed within the Milky Way and nearby galaxies, probed primarily with integrated intensity or column density images, can be adequately modeled with a single power-law component
\begin{equation}\label{eq:single_comp_SPS}
    P\left(k\right) = A k^{\gamma}+B, 
\end{equation}
where $A$ is the power-law amplitude, $\gamma$ is the power-law index, and $B$ describes the contribution from peculiarities within the image such as point-sources or instrumental noise. However, several studies of external galaxies (e.g., \citealt{Elmegreen2001, Dutta2009, Combes2012}) find power spectra are best modeled by broken power laws, indicating a difference in either the physical processes dominating at particular scales \citep{swift2008}, or influence from observation effects \citep{koch2020}. To capture this behavior, an estimated break scale can be passed to the fitting routines in {\tt Turbustat} to enable fitting the power as a function of scale with a segmented linear model that iteratively optimizes the location of the break point. An optimal break point minimizes the difference between the different model parameters while also partitioning the power into two distinct power-law relationships defined as
\begin{equation}\label{eq:two_comp_SPS}
     P\left(k\right) = 
        \begin{cases} 
          A k^{\gamma_{\rm large}} & k\leq k_{\rm b} \\
          B k^{\gamma_{\rm small}} & k >k_{\rm b}  \\
       \end{cases},
\end{equation}
where the power law indices and amplitudes are now partitioned to represent large and small scales on either side of the break scale, $k_{\rm b}$. The fitting routine reverts to a single component power-law fit if no optimal break point is found. An additional term, $C$, is included to capture power from point-like sources.

Previous studies (e.g., \citealt{koch2020}) have defined custom models and fitting routines after constructing the azimuthally-averaged  1D power spectrum profile of astronomical images in order to account for instrumental systematics such as the instrumental point spread function (PSF) response and noise. Visual inspection of the fits to the various timesteps and phase maps (e.g., Figure~\ref{fig:SPS}) demonstrate that the default fitting routines available in {\tt Turbustat} sufficiently model the measured power spectra and reliably determine the break scale in the absence of  typical observational systematics. For each phase and simulated time step, we provide the {\tt PowerSpectrum} fitting routines with an initial estimate of the break scale equal to 600 pc. Our results are not changed by reasonable variations from this initial estimate for the break scale. 

\subsection{\bk{Analysis period}}
\bk{Our analysis starts at $t=262\,\mathrm{Myr}$, which is about one rotation 
of the galaxy at the solar distance. This time also marks the onset of disc 
fragmentation due to the Parker instability \citep[i.e.][]{Koertgen18L}. 
Before this time, the disc density profile is rather smooth and the overall 
dynamics are dominated by the rotation. In contrast, from this chosen time on, 
the disc gravitational and Parker instability self-consistently generate a dynamical ISM in 
the galaxy. The timesteps for our 
analysis are separated by $\Delta t=50\,\mathrm{Myr}$, which is about 20\,\% 
of an orbit at solar distance to the center and ensures that the dynamics 
over a full rotation are sufficiently time-resolved. At the same time, the time interval 
is, although slightly larger, comparable to the observed typical lifetime of 
molecular clouds in external galaxies \citep{Chevance2020} and thus provides 
insight into the dynamical ISM within a lifetime of a molecular cloud. In total, we thus encompass almost two full orbits at solar distance with our analysis and capture small-scale as well as large-scale, galactic dynamical processes.}

\section{Results}\label{sec:results}
\subsection{\bk{General Overview}}
\begin{figure*}
    \centering
        \rotatebox{90}{\quad\quad\quad\quad \bk{$t=262\,\mathrm{Myr}$}}\includegraphics[width=0.27\textwidth]{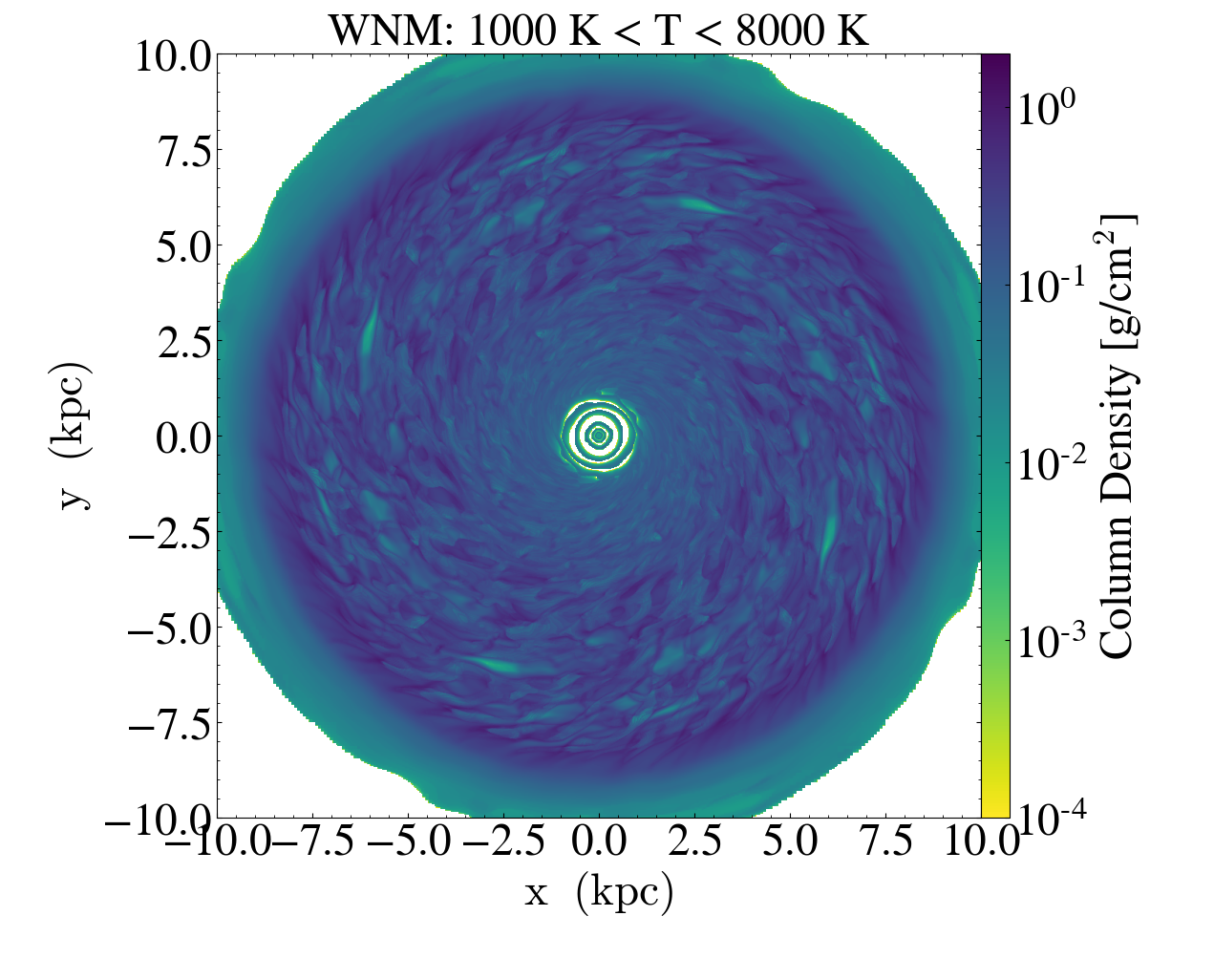}\includegraphics[width=0.27\textwidth]{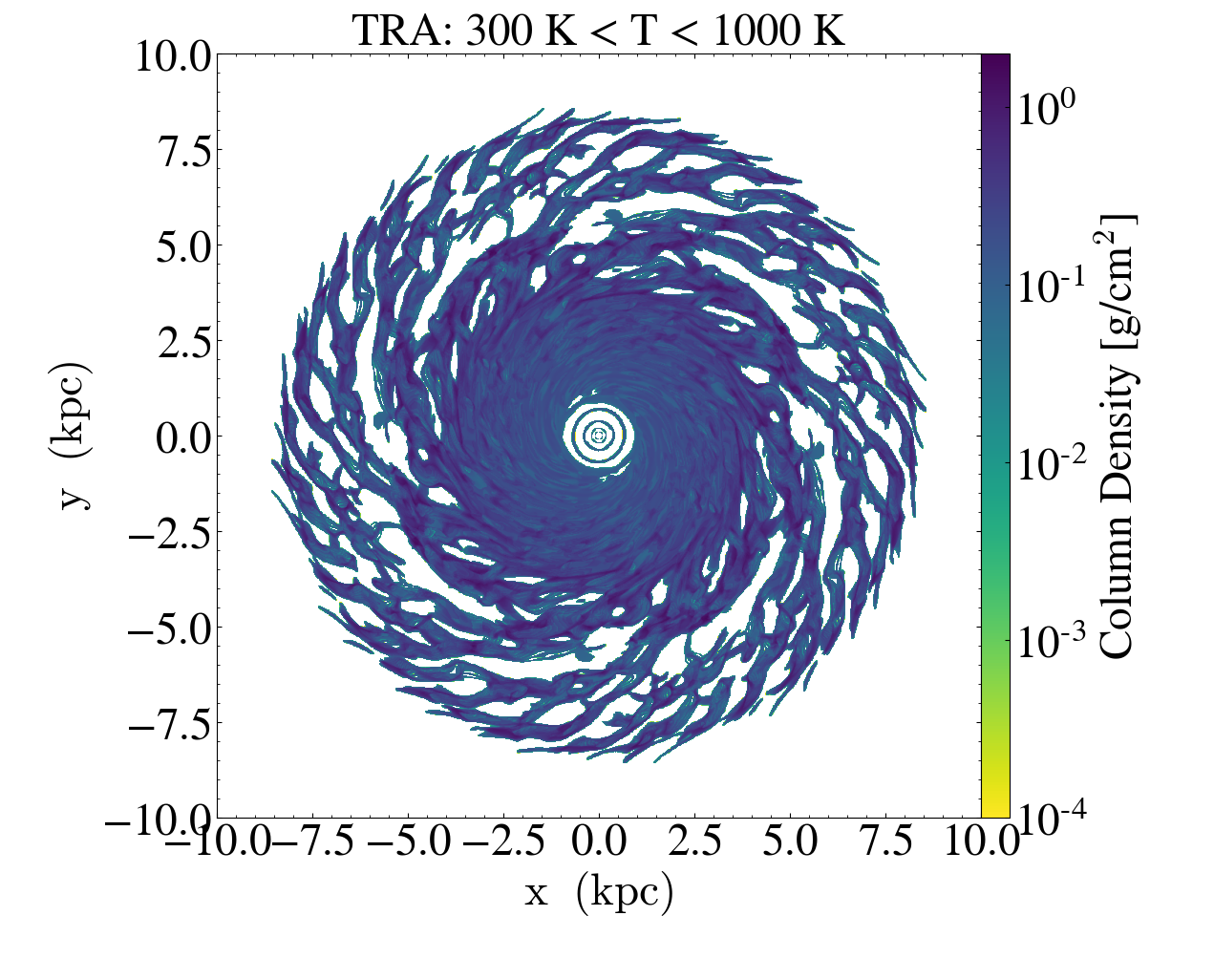}\includegraphics[width=0.27\textwidth]{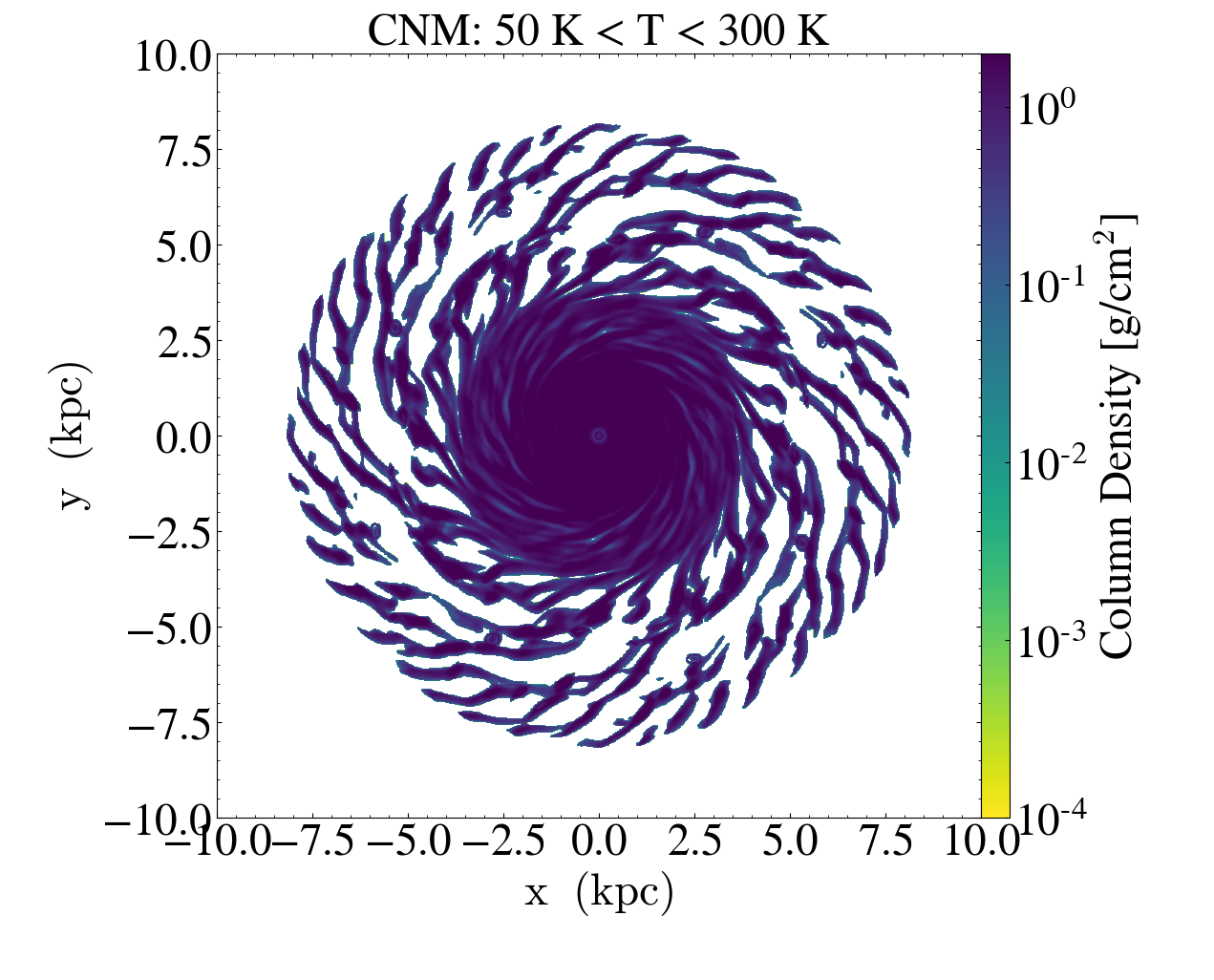}\includegraphics[width=0.27\textwidth]{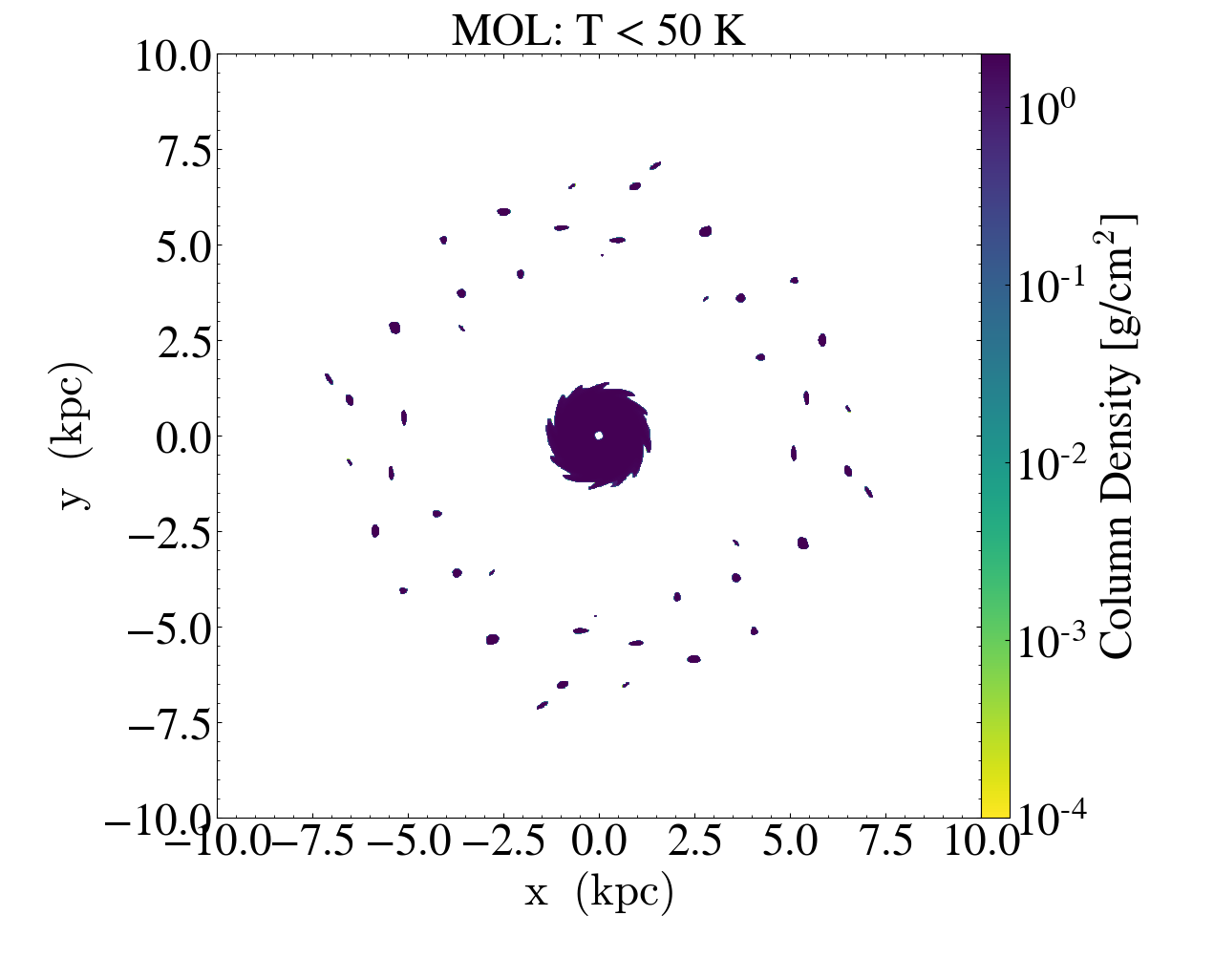}\\
        \rotatebox{90}{\quad\quad\quad\quad \bk{$t=462\,\mathrm{Myr}$}}\includegraphics[width=0.27\textwidth]{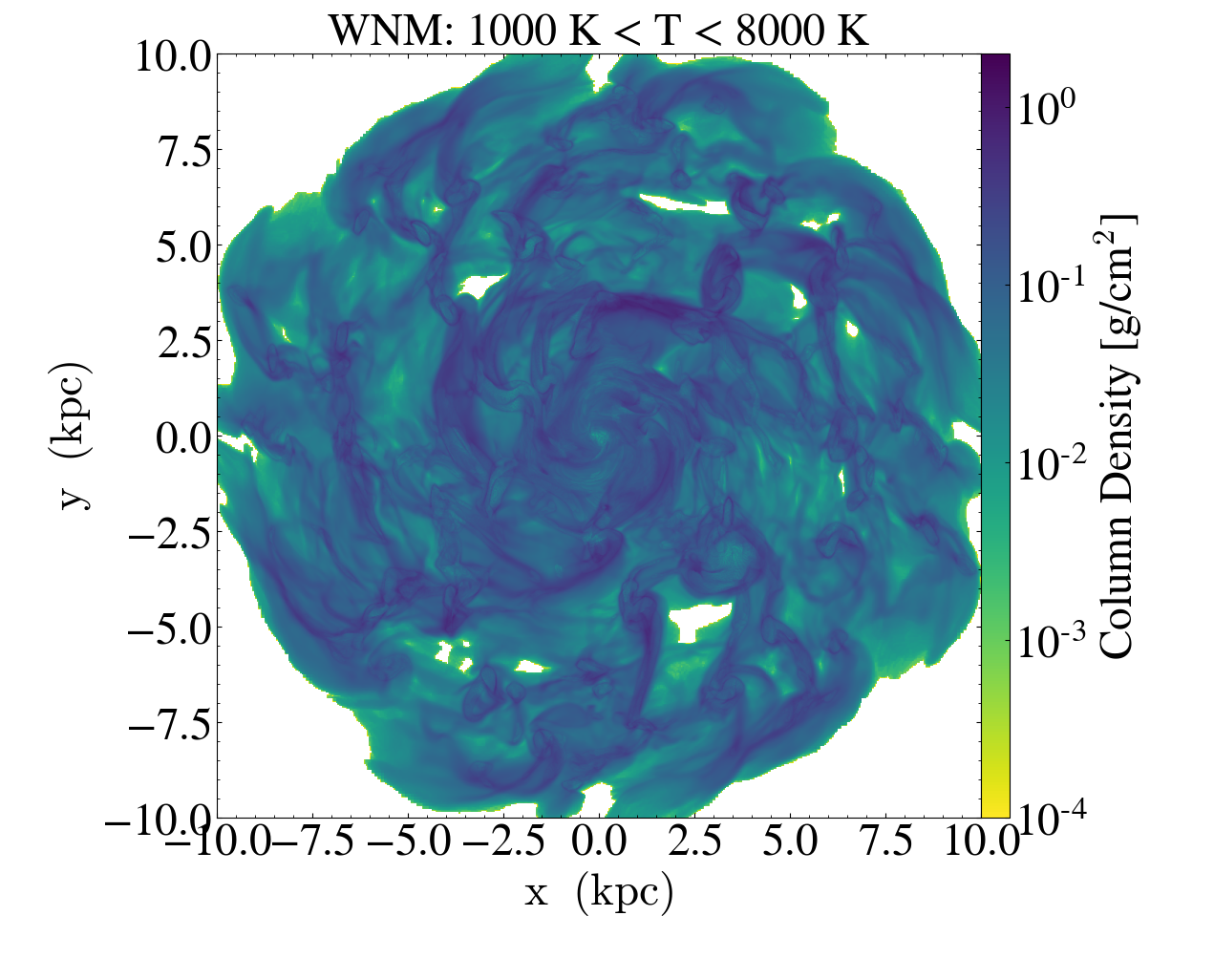}\includegraphics[width=0.27\textwidth]{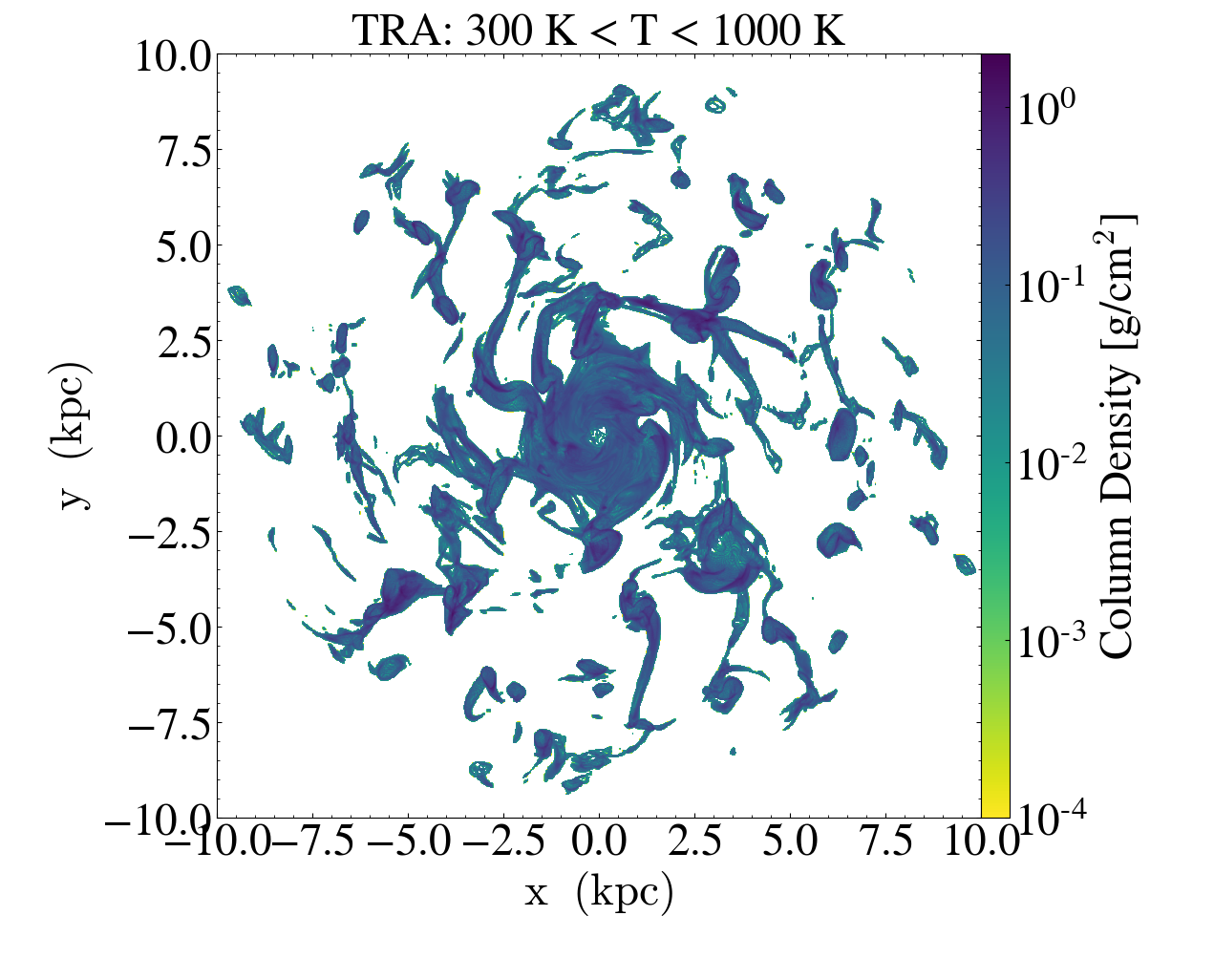}\includegraphics[width=0.27\textwidth]{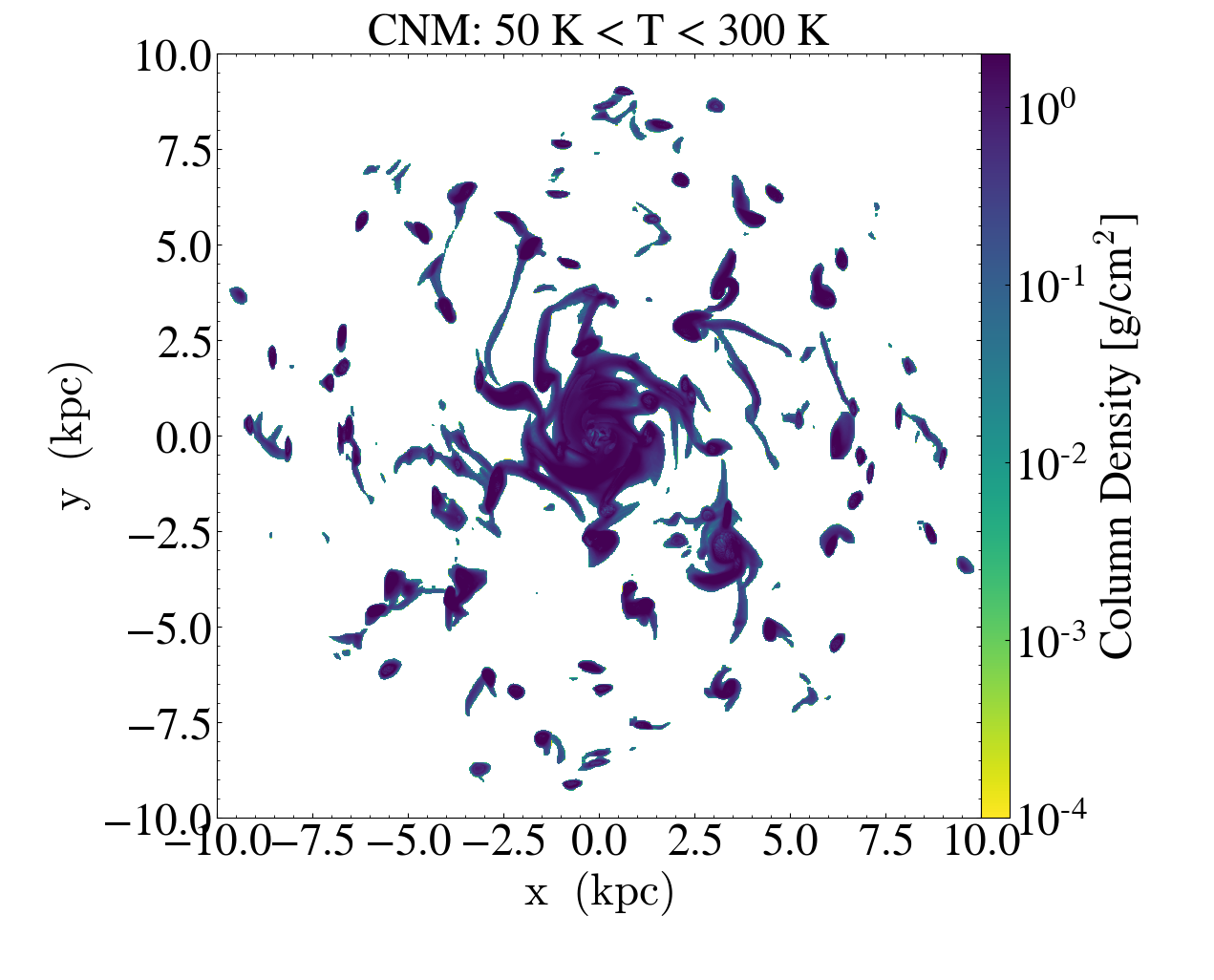}\includegraphics[width=0.27\textwidth]{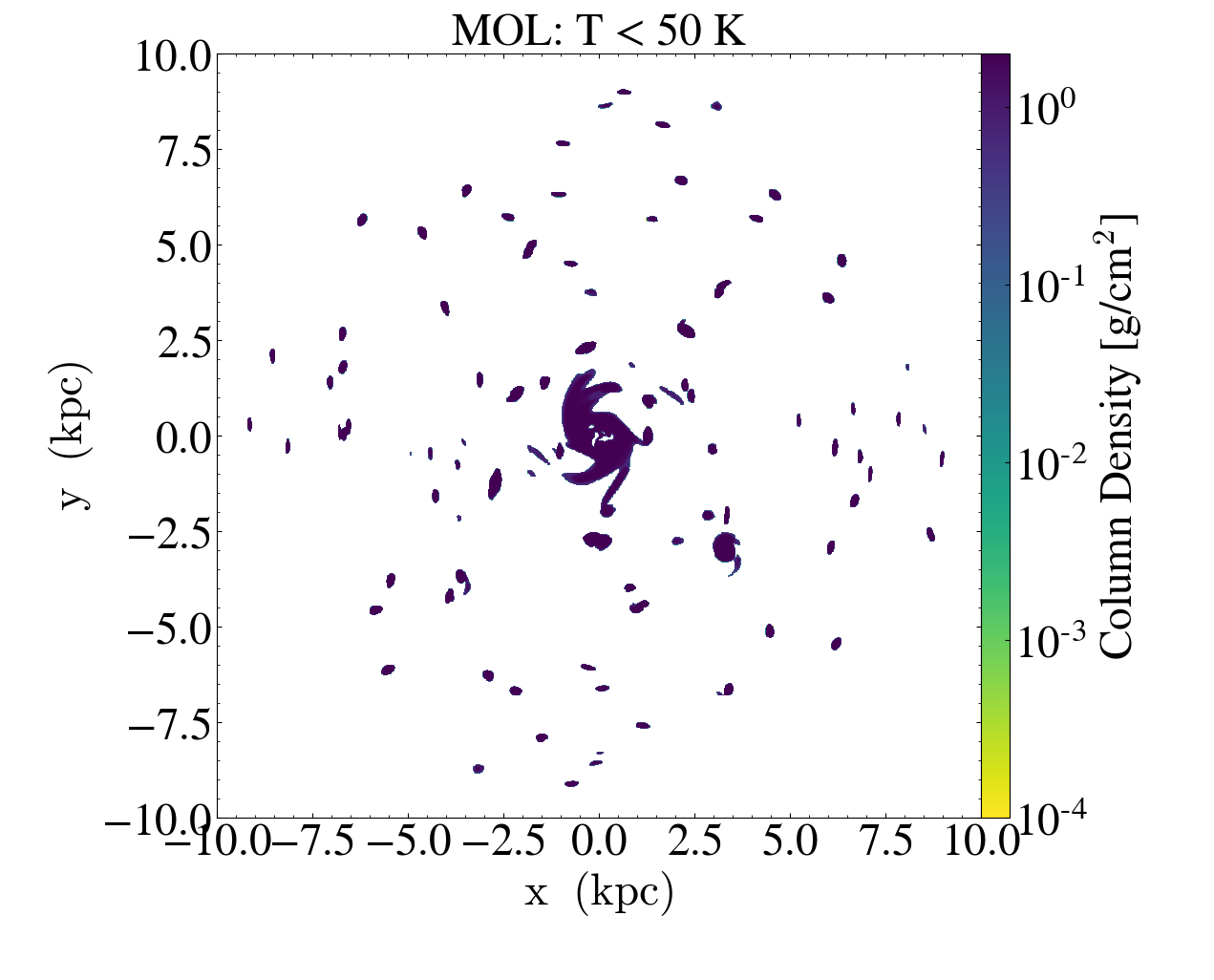}\\
        \rotatebox{90}{\quad\quad\quad\quad \bk{$t=662\,\mathrm{Myr}$}}\includegraphics[width=0.27\textwidth]{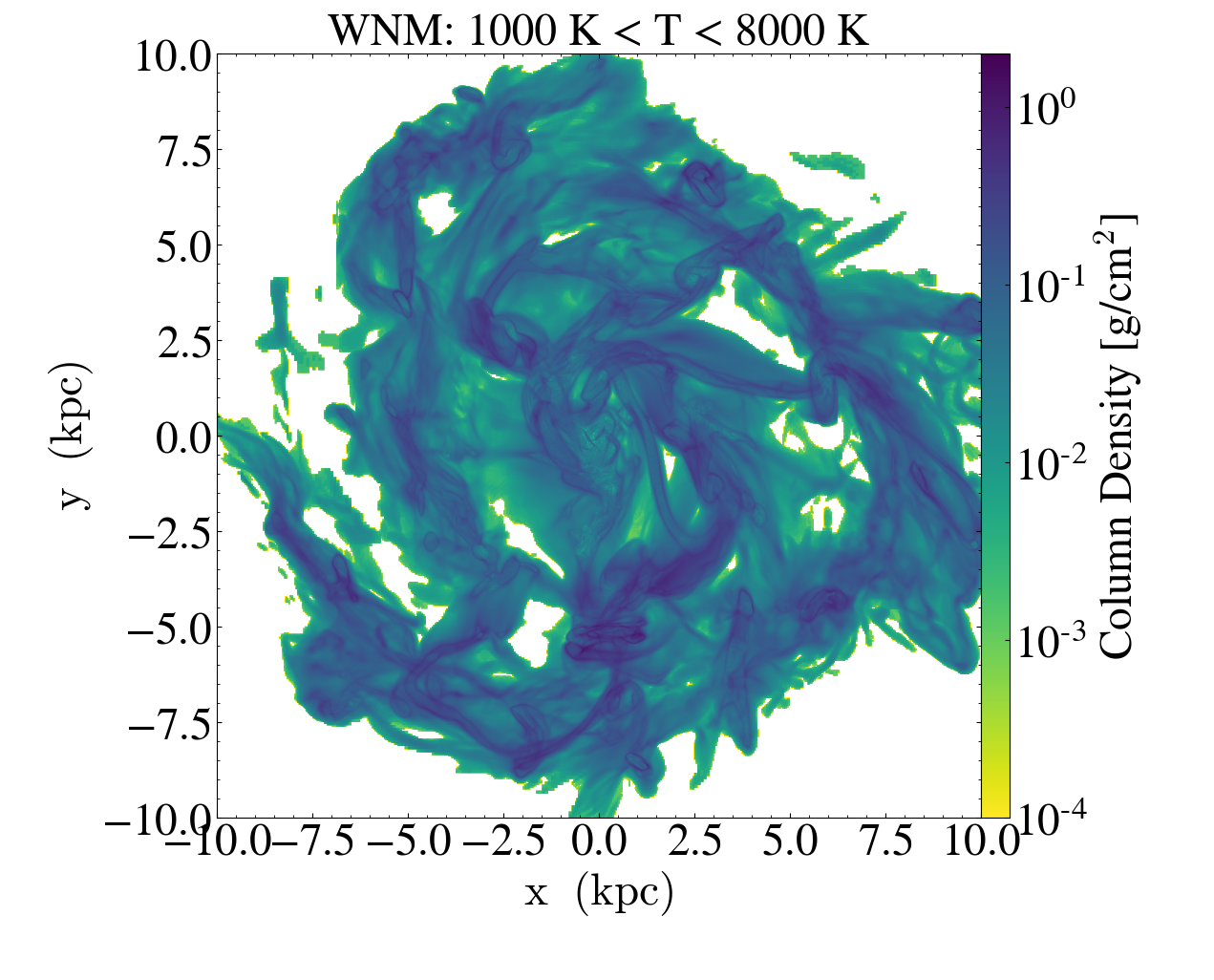}\includegraphics[width=0.27\textwidth]{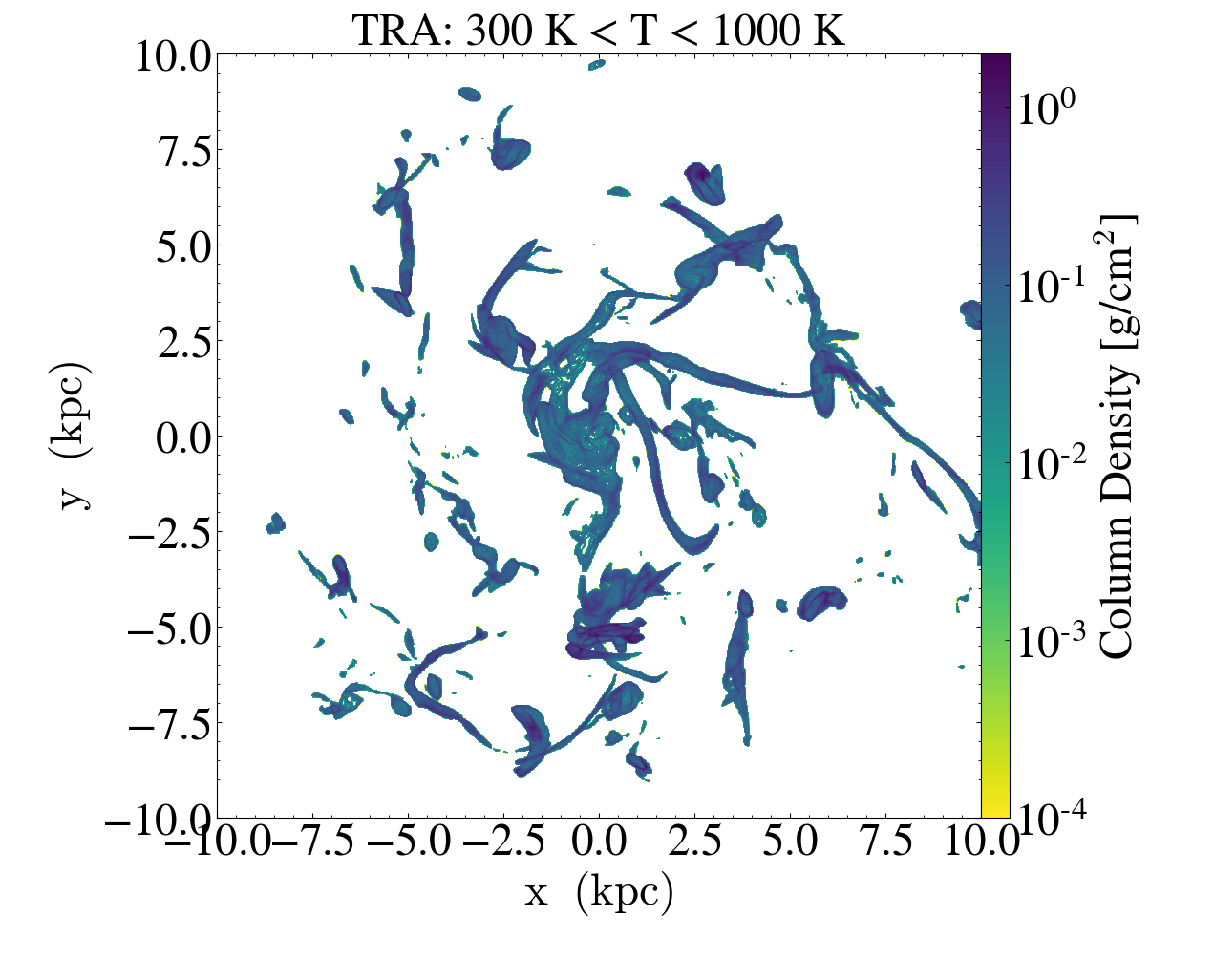}\includegraphics[width=0.27\textwidth]{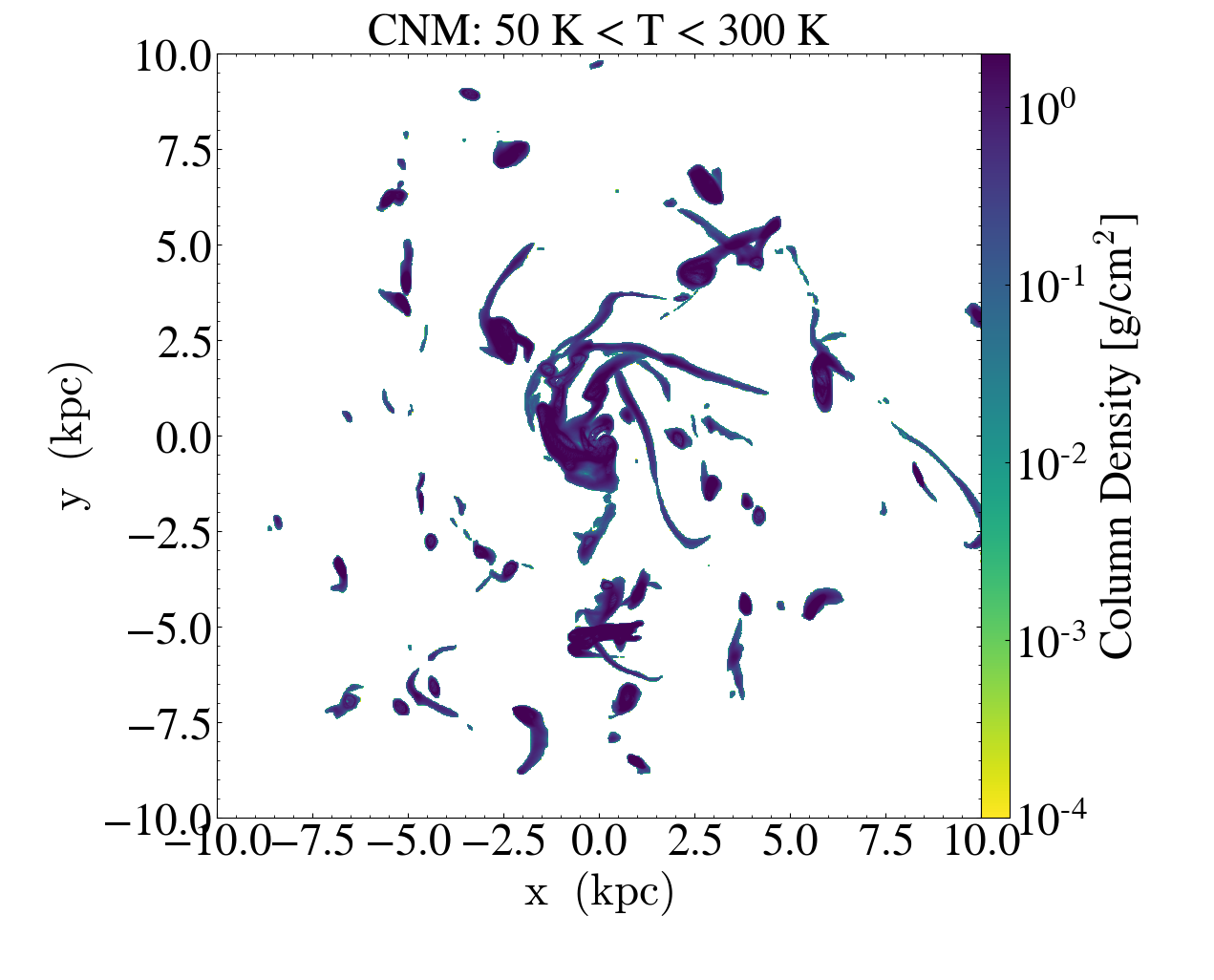}\includegraphics[width=0.27\textwidth]{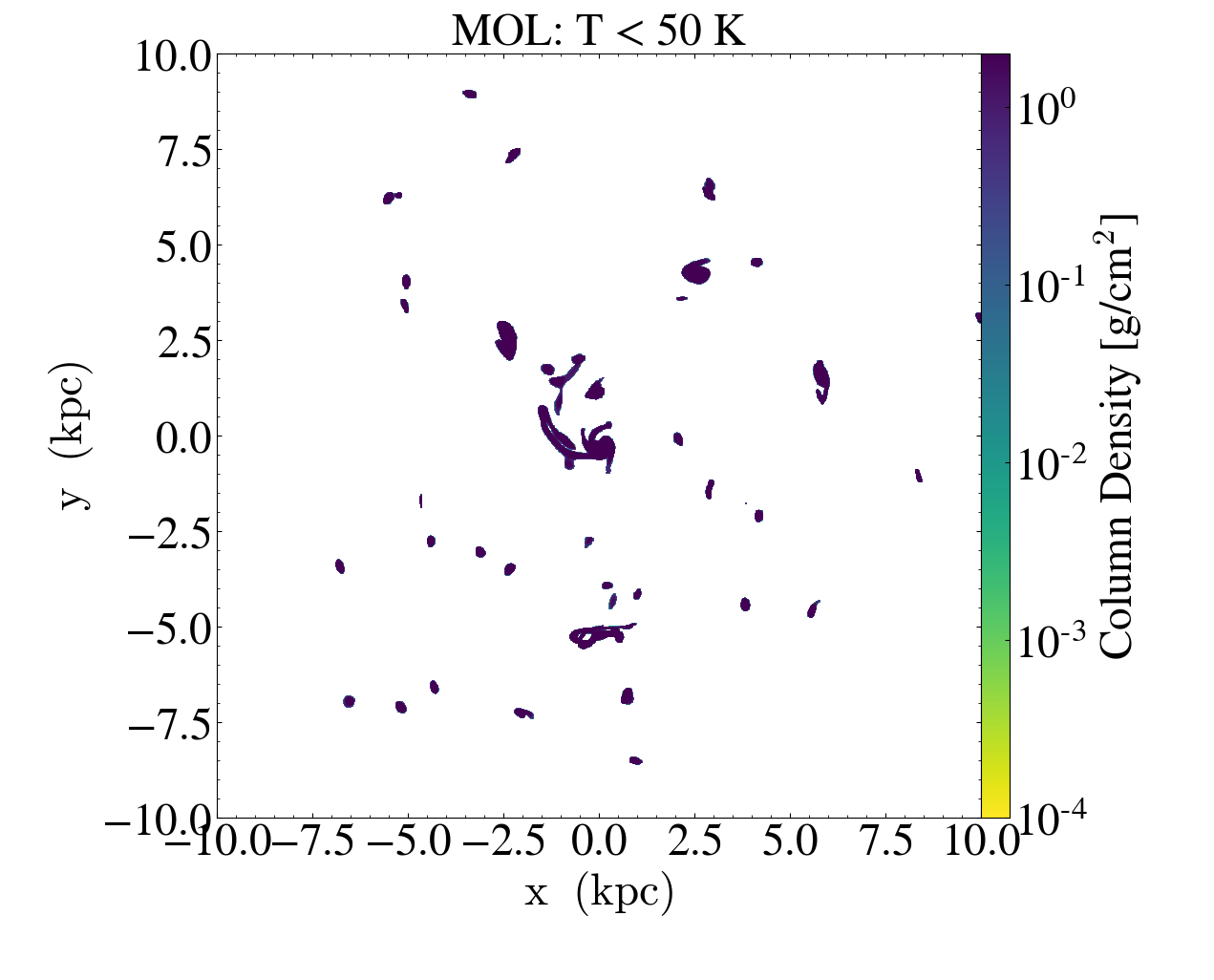}
        \caption{Column density maps of four phases. \ita{Left to right:} Warm \Hone (WNM), transitional gas (TRA), cold \Hone (CNM) and H$_2$ gas (\bk{MOL}). \ita{Top to bottom:} Time evolution in steps of 200\,Myr, corresponding to almost an entire orbit at 
        $R=8\,\mathrm{kpc}$. The corresponding temperature range is indicated in each figure title. Note 
        the filamentary morphology of the gas structures in all phases.}
        \label{fig:maps}
\end{figure*}

\begin{figure*}
    \centering
    \includegraphics[width=0.5\textwidth]{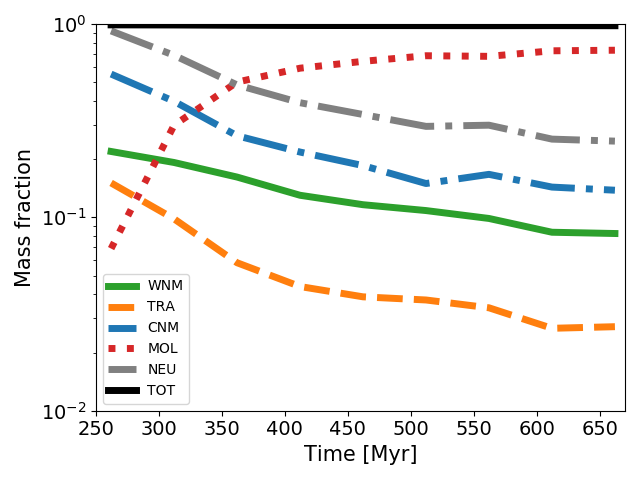}\includegraphics[width=0.5\textwidth]{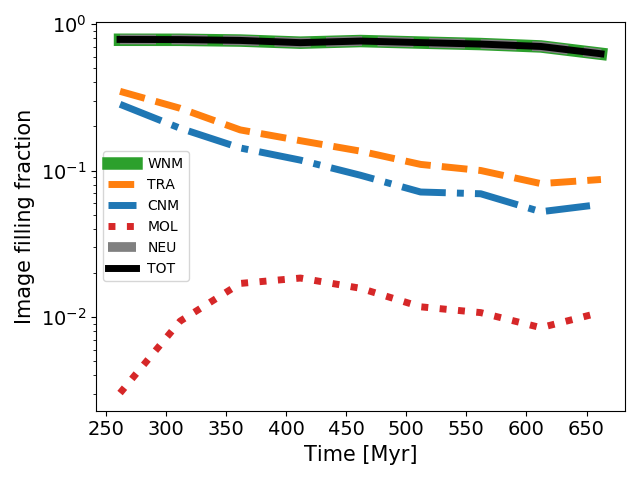}
    \caption{\bk{Time evolution of the mass (left) and image area filling (right) fractions. The mass fraction is determined in a 3D cylindrical volume, which contains the entire 
    galaxy. The image filling fraction is determined for an image with edge length of 
    20\,kpc. Due to the absence of stellar feedback the fraction of gas in the warmer, diffuse phases decreases, while the amount of gas in the densest and coldest phase 
    continuously increases. However, as can be seen from the right panel, most of the mass 
    resides in regions that only fill about a percent of the total area, while the WNM 
    almost fills the entire image extent. 
    Note that the NEU phase is hidden by the 
    TOT phase in the right panel.}}
    \label{fig:mfaff}
\end{figure*}

In Fig.~\ref{fig:maps} we show column density maps of the galaxy at three times and for the various gas 
phases. The first row shows the galaxy at $t=262\,$Myr. The second and third row show maps of the 
galaxy at $t=462\,$Myr and $t=662\,$Myr, respectively.\\
At the earliest stage shown, the WNM gas is rather smoothly distributed over the galaxy, but shows some 
flocculent structure locally. Most of the WNM gas up to $R\sim8\,\mathrm{kpc}$ shows column densities in the 
range of a few $\left(10^{-2}-10^{-1}\right)\,\mathrm{g\,cm}^{-2}$. The transitional phase shows a clear 
filamentary structure, which was imprinted by the disc fragmentation process. The white areas in between 
correspond to warmer gas structures. The CNM is observed to be even more filamentary and also much 
denser. Most filaments are observed to be 1-2 kpc long. As can be further seen from the \bk{MOL} map, some 
filaments of the CNM already harbour colder and denser clouds. The innermost part of the galaxy has not 
fragmented yet, as is inferred from the smooth patch in the central part of the cold gas maps.\\
With time, galaxy dynamics produce turbulent motions in the galaxy, which tend to disrupt the smooth 
density structures. We also see the innermost part of the disc fragmenting into filamentary structures and the morphology 
of the structures in the galaxy has become highly time dependent. We emphasize that not all CNM structures 
further cool down and fragment into \bk{even denser} structures.\\
\bk{In Fig.~\ref{fig:mfaff} we additionally show the time evolution of the mass fraction 
and the image filling fraction of the various phases. As expected, most of the mass is in 
the MOL phase with a mass fraction of almost 70\,\%. The remainder is dominated by the CNM 
phase as the coldest of the three additional phases. By the end of the simulation, the 
WNM reveals a mass fraction of about 10\,\%. The TRA phase shows the lowest mass fraction 
due to it representing the thermally unstable regime.\\
For the image filling fraction, the picture is 
contrary to the evolution of the mass fraction. Whereas the mass budget is dominated by 
MOL phase, it clearly only fills around a percent of the image. Hence, most of the mass 
in the galaxy is locked up in very small, dense patches, i.e. molecular clouds. The WNM 
fills almost the entire image and, thus, reveals a wide spread distribution. Lastly, the 
combined phases NEU and TOT reveal the 
expected behaviour, i.e. the TOT phase sums 
up to about $97-99$\,\% of the mass in the 
disc, indicating that there is still some very small portion of gas warmer than $T=8000$\,K. 
For the image filling fraction, these two 
phases are indistinguishable from the WNM 
phase.}

\subsection{The Spatial Power Spectrum and the Break Scale}
\begin{figure}
    \centering
        \includegraphics[width=0.48\textwidth]{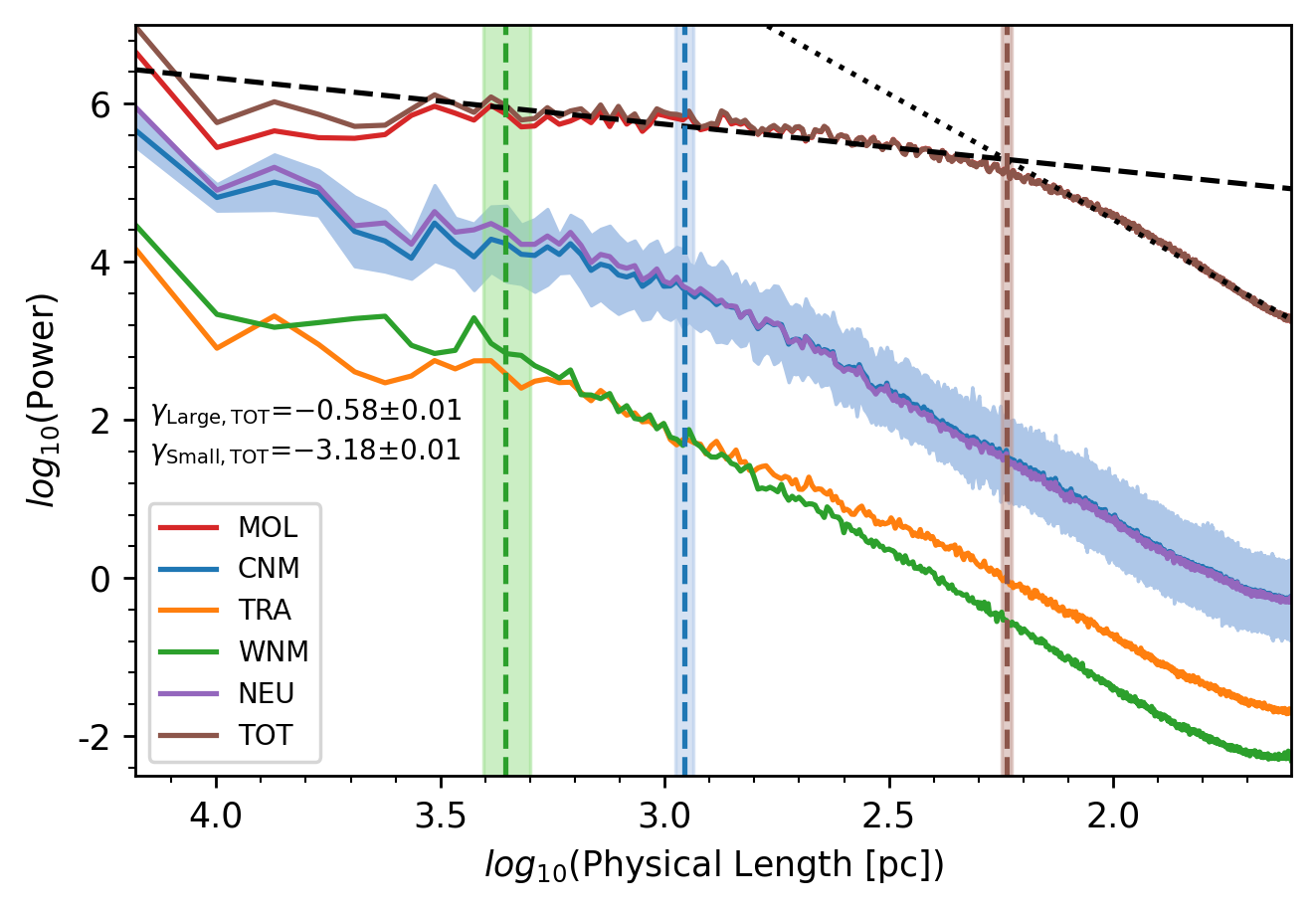}
        \caption{Exemplary spatial power spectrum for various 
        phases \bkk{at $t=450\,\mathrm{Myr}$}. It is clear that the SPS of the total gas 
        is dominated by the \np{MOL} phase, which contributes 
        most to the column density. The large- and small-scale slopes, as well as the break scale, are indicated for the total \np{TOT} phase (\np{MOL}+CNM+WNM+TRA) \np{and NEU phase (CNM+WNM+TRA)}. The vertical dashed lines mark the loci of the break scales for the warm \Hone\np{, CNM, and} the total \np{neutral} phase. The shading on the CNM phase gives the respective uncertainty and is representative of the uncertainty for the other phases. 
        Converting the power spectrum to an energy spectrum 
        by multiplying by $k^{D_i-1}$, where $D_i$ is the 
        number of dimensions of the input data, that is $D_i=2$, the slopes become \np{$\gamma_\mathrm{Large}=0.58$} and \np{$\gamma_\mathrm{Small}=-2.18$}. Hence, the SPS 
        implies shock-dominated turbulence on scales 
        below the break scale.}
        \label{fig:SPS}
\end{figure}

We produce azimuthally-binned SPS profiles of the 2D surface density (i.e., column density) images for each phase and several time steps. As an example, we show in Fig.~\ref{fig:SPS} the SPS for the \np{MOL}, WNM, TRA and CNM phases, as well as of their 
combined signal (\bk{NEU and} TOT). The power in each spectrum is essentially proportional to the respective 
column density, so that the colder (and denser) phase dominates in power. {\tt Turbustat} identifies a 
break in the spectrum, for which we highlight its corresponding scale by the vertical dashed lines. The break also implies a change in slope of the spectrum. For the TOT phase, the slope is flatter on scales larger than the break and becomes 
steeper at smaller scales. The slope changes by \bk{$\left|\Delta\gamma\right|\sim2.5$} in this example. The 
vertical lines further emphasize that the break scale is phase dependent, owing to the varying 
morphology of structure in the maps and possibly to varying dynamics in the different phases. 

\subsubsection{Time Evolution of the SPS Properties}
\begin{figure}
    \centering
    \includegraphics[width=0.5\textwidth]{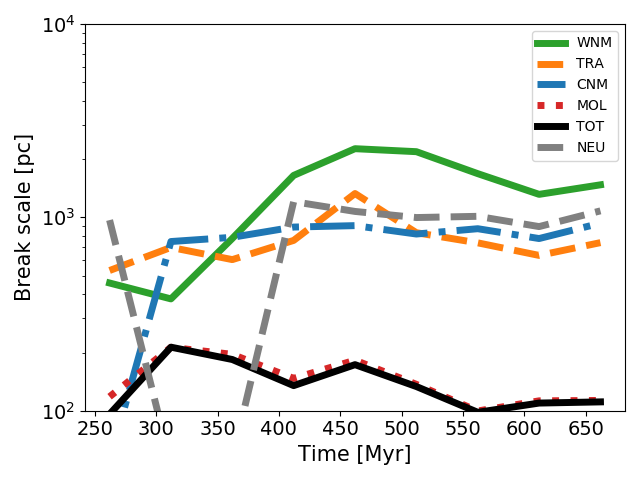}
    \caption{Time evolution of the break scale for the various phases. The behaviour of the break scale 
    clearly shows a phase-dependence. While it increases over time for the warm \Hone phase, it slightly 
    decreases for the H$_2$ phase. In contrast, the intermediate temperature phases show only few variations, and the break scale is quite similar.}
    \label{fig:breakscale}
\end{figure}

\begin{figure*}
    \centering
    \includegraphics[width=0.5\textwidth]{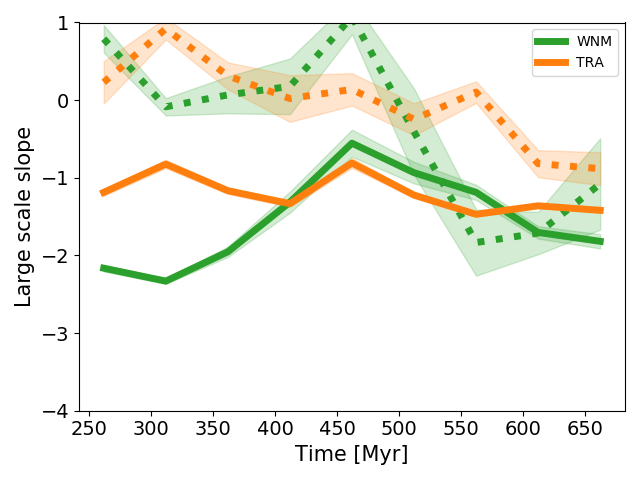}\includegraphics[width=0.5\textwidth]{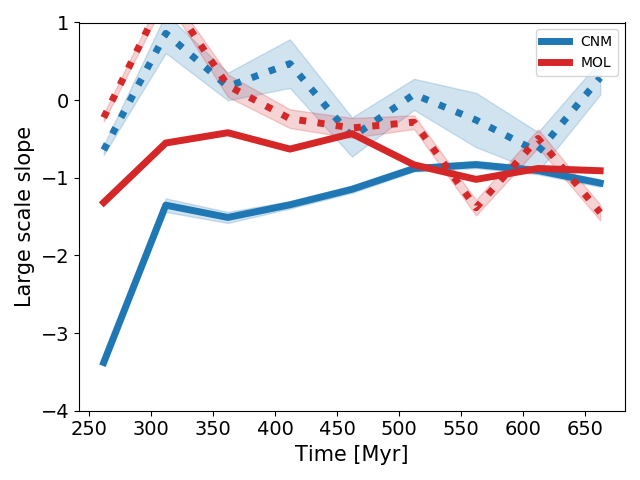}\\
    \includegraphics[width=0.5\textwidth]{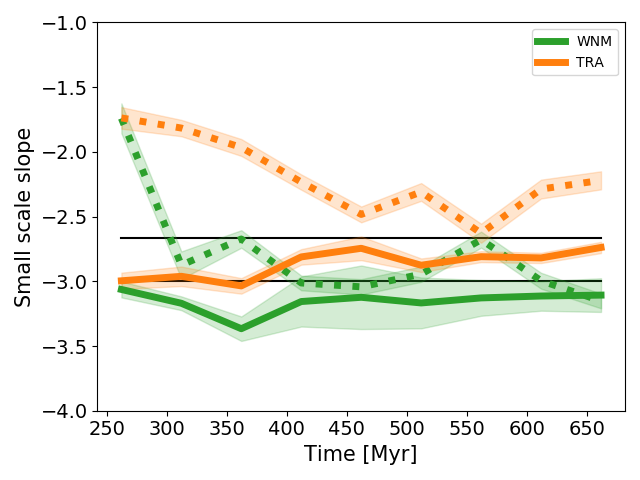}\includegraphics[width=0.5\textwidth]{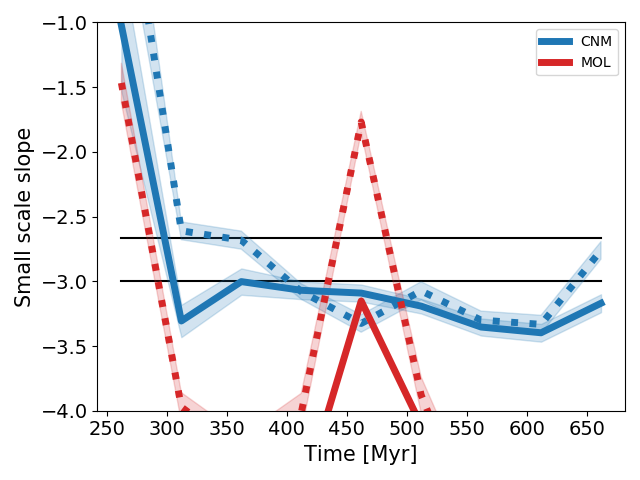}\\
    \caption{Time evolution of the large- (top) and 
    small-scale (bottom) slopes of the SPS (solid lines) and the kinetic energy power spectra (dotted lines) for the various 
    phases. Black horizontal lines indicate the power spectrum slopes of 
    Burgers and Kolmogorov turbulence at $\gamma =$ -3 and -8/3 respectively. The data indicate that at late times, the 
    energy spectra are essentially flat above the break 
    scale and resemble almost shock-dominated / 
    Burgers type slopes below it. Please note 
    that, at early times, the shallow small-scale slopes of 
    the CNM and H$_2$ phases are due to ongoing 
    fragmentation of the disc and subsequent 
    condensation of cool, dense gas out of the 
    warm and diffuse one. At later times a full 
    cascade has emerged. Further note the highly 
    fluctuating slope of the H$_2$ phase, which 
    we primarily attribute to its small volume 
    filling fraction.}
    \label{fig:slopes}
\end{figure*}

To further characterize the SPS, we present the time evolution of the break scale for the 
phases in Fig.~\ref{fig:breakscale}. It is obvious that the break scale is phase dependent 
and further shows a small variation with time. The break in the SPS of the WNM moves to 
larger scales with time, while the two phases TRA and CNM 
do not show significant time variation and further appear quite similar in value. In 
the latter two cases, the respective phases are replenished from the warmer phases due to gas 
cooling (and probably shock heating). The SPS of these phases are thus the least affected 
by the overall dynamics. The densest and coldest
phase of our analysis, H$_2$, shows a break scale that ranges between 100 and 200 pc, with 
a slight decrease over time. This phase is mainly composed of dense clouds, which 
resemble the small scale structures within our model galaxy.\\
\np{Both NEU and TOT lines show that}, \bk{in agreement with 
Fig.~\ref{fig:SPS}, it is indeed the densest phase, which dominates the SPS of galaxies, 
although the break scale in the NEU phase shows some contribution from the TRA and WNM 
gas.\\ }
The SPS is further characterized by the slopes of the spectrum above and below the 
break. Fig.~\ref{fig:slopes} shows the time evolution of the large (above the break) and 
small scale (below the break) slopes for the various phases as solid lines. The large scale slopes show values around $-$1 for all phases with some temporal evolution. While the slopes for the TRA and H$_2$ phase seem to steepen slightly (getting more negative), the slopes for the WNM and CNM first flatten and, in case of the WNM, steepen again. In contrast, the small scale slopes show even less variation for the WNM, TRA and CNM gas. The latter two further support the conclusion from above that the SPS of the TRA and CNM phases does 
not vary too much. Typical values of the small scale slopes are around $-$3.2 for the WNM and CNM and between -2.67 and -3 for the TRA phase. The large variation of the H$_2$ slope is due 
to the small volume filling factor of the gas structures.\\
The energy spectrum is related to the power spectrum via 
\beq
E(k)\propto k^{D_\mathrm{i}-1}P(k),
\eeq
with the number of input dimensions $D_\mathrm{i}$, in our case \mbox{$D_\mathrm{i}=2$}. 
Incompressible (Kolmogorov) and compressive (Burgers) turbulence have energy spectra of 
the form $E_K(k)\propto k^{-5/3}$ and $E_B(k)\propto k^{-2}$, respectively. The corresponding slopes of the power spectra, $-8/3$ and $-3$, are indicated in the panels as black solid lines. Under the assumption that the density field is treated as a passive scalar and the dynamics of the velocity field are imprinted in the density field, the turbulence in the phases is of Burgers type, typical for supersonic motions. The TRA phase, however, shows some 
trend from shock dominated (Burgers) to incompressible Kolmogorov turbulence.\\
The dotted lines in Fig.~\ref{fig:slopes} show the large and small scale slopes of the 
corresponding kinetic energy power spectra ($\Sigma^{1/2}v_\mathrm{los}$). Initially, 
the large scale slopes are positive, indicating that the break scale resembles the scale of maximum energy (injection). With time, however, they decrease and become negative. This implies that the scale of maximum energy injection moves to scales larger than the 
disc size. The large difference between the slopes of the SPS and the kinetic energy power spectra at early times is in parts due to small velocities along the line of sight, as we start with an entirely toroidal velocity field. At late times, the slopes of the density SPS and kinetic energy spectra become more similar, 
which indicates the coupling between these fields.

\section{Discussion}\label{sec:discussion}
\subsection{Origin of the Break Scale}
\begin{figure*}
    \begin{tabular}{cc}
    \Large{\fat{WNM}}   &\Large{\fat{\bk{MOL}}}\\
    \includegraphics[width=0.5\textwidth]{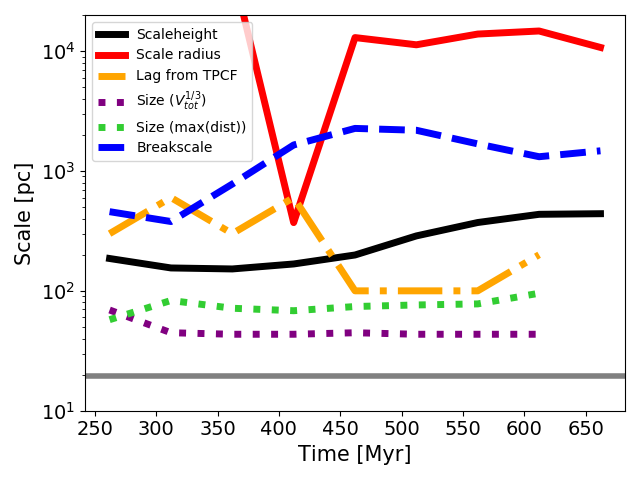}&\includegraphics[width=0.5\textwidth]{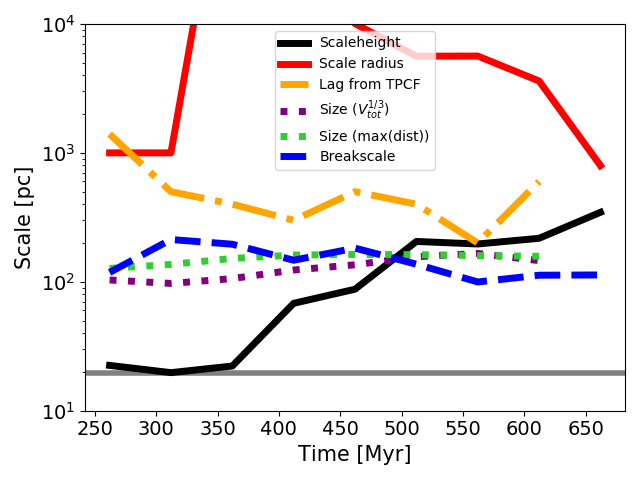}
    \end{tabular}
    \caption{Comparison of the various derived scales with 
    the break scale from the SPS. While the break scale of the warm phase (left) does not correspond to any 
    characteristic scale of the galaxy, it strikingly resembles the typical size of objects in the 
    face-on maps for the \bk{MOL} gas (right). Interestingly, the break scale does \ita{not} match the 
    disc scale height in either phase. The horizontal grey solid line indicates the cell size.}
    \label{fig:scales}
\end{figure*}

\begin{figure*}
    \begin{tabular}{ccc}
    \fat{WNM}   &\fat{CNM}  &\fat{\bk{MOL}}\\
    \includegraphics[width=0.33\textwidth]{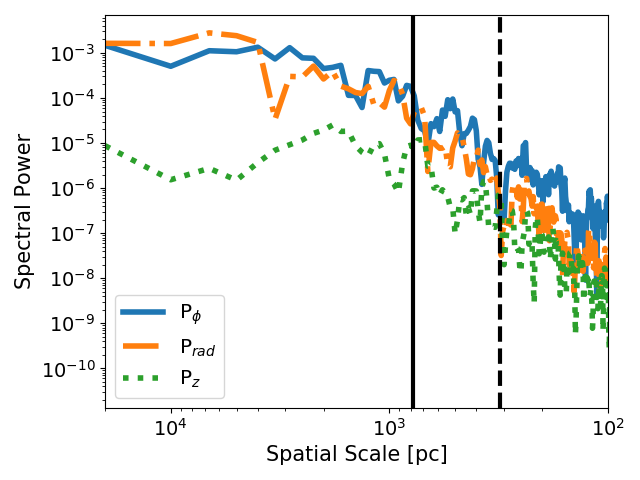}& \includegraphics[width=0.33\textwidth]{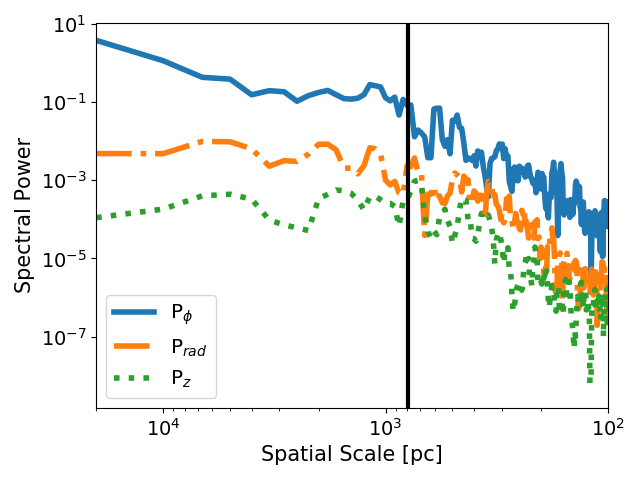}& \includegraphics[width=0.33\textwidth]{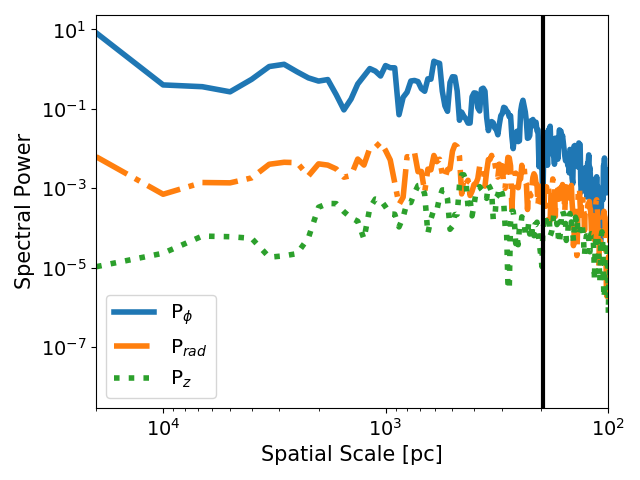}\\
    \includegraphics[width=0.33\textwidth]{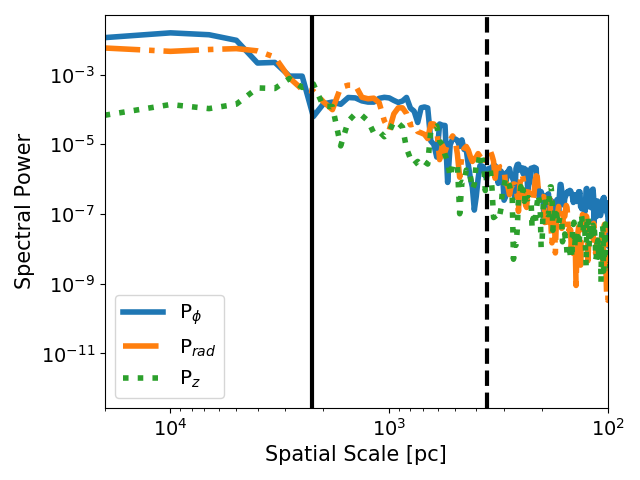}& \includegraphics[width=0.33\textwidth]{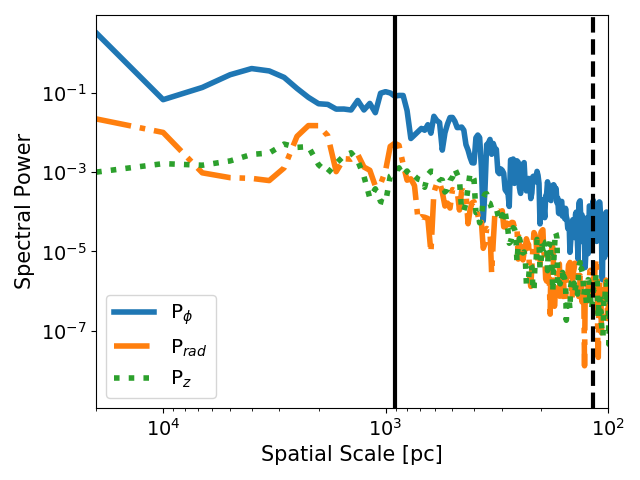}& \includegraphics[width=0.33\textwidth]{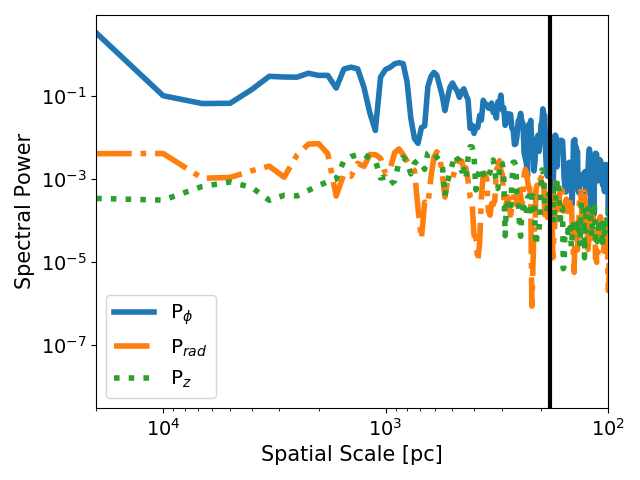}\\
    \includegraphics[width=0.33\textwidth]{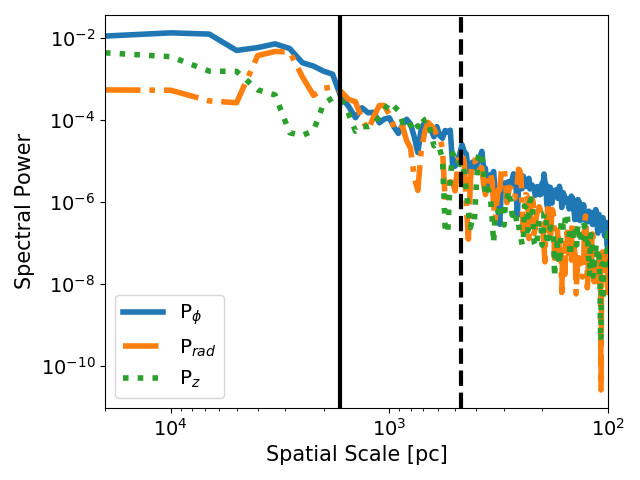}& \includegraphics[width=0.33\textwidth]{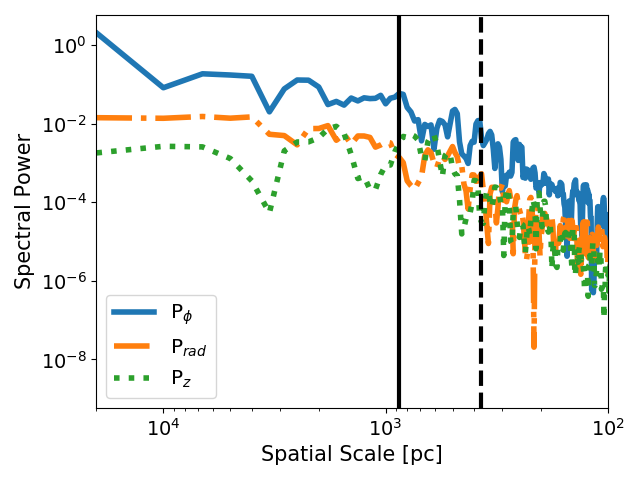}& \includegraphics[width=0.33\textwidth]{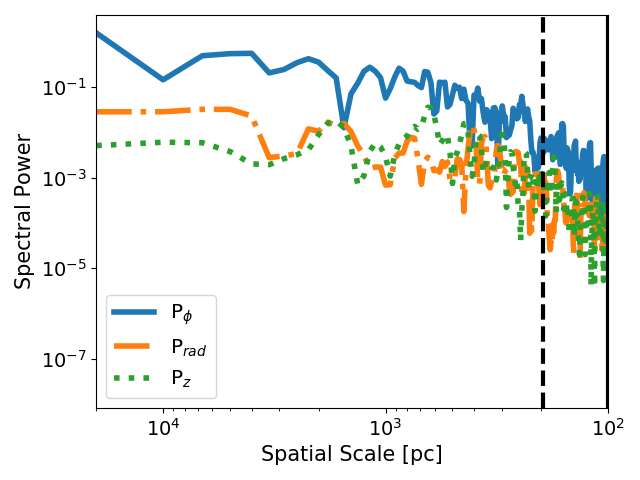}\\
    \includegraphics[width=0.33\textwidth]{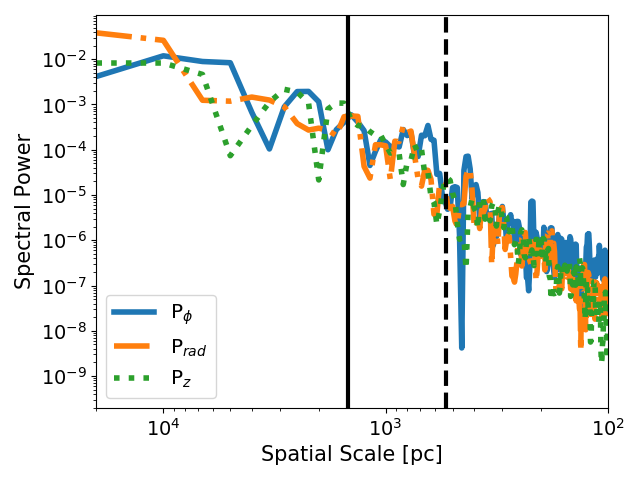}& \includegraphics[width=0.33\textwidth]{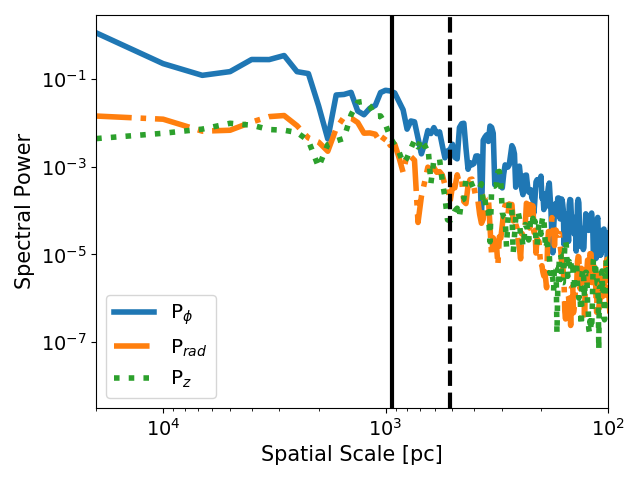}& \includegraphics[width=0.33\textwidth]{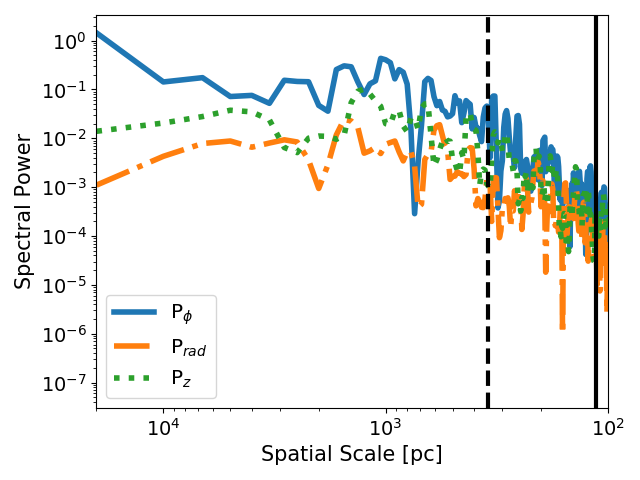}\\
    \end{tabular}
    \caption{Kinetic energy power spectra of three phases (left to right) for various times 
    (top: early, bottom: late). All velocities are first projected along the line of sight parallel 
    to the rotation axis of the galaxy. The bulk rotational motion of the galaxy is 
    subtracted for annuli in bins of $\Delta R=100\,\mathrm{pc}$. The black vertical lines mark 
    the break scale as estimated from the spatial power spectrum (solid), as well as the 
    scale height of the respective phase estimated by projecting the disc edge-on and 
    fitting the resulting column density profile (dashed). While the break scale nicely traces the 
    size of the point sources of the H$_2$ phase, it fails to describe the warm \Hone gas 
    structures. This is primarily due to the smoother density maps. However, we can see in this figure the break 
    scale instead traces a transition in the turbulent regime from a more \bk{disc-like} 
    one above the break to a rather \bk{3D} behaviour below it.}
    \label{fig:spectra}
\end{figure*}

A break in the SPS has been found in many studies \citep[e.g.,][]{Elmegreen2001,Dutta2008}. As elaborated above, this break has been 
interpreted as the disc scale height both in observations and numerical simulations \citep{Bournaud2010}. In contrast, a recent study  by \citet[][\bk{see also numerical indications by \citet{Grisdale2017}}]{koch2020} questions 
these previous findings and claims instead that the observed break was due to the PSF of the telescopes. They were able to show through forward modeling of the PSF that the breaks in the SPS can be attributed to the decrease in the PSF response. They also argue, the distribution of point sources, that is, molecular clouds, may give rise to a characteristic scale in the SPS.\\
In Fig.~\ref{fig:scales} we show the time evolution of various derived scales and compare them 
with the break scale in the SPS for the WNM (left) and \bk{MOL} phase (right). We determine the disc scale height by generating an averaged edge-on column density profile and fit this with a $sech$ function. In case of the WNM phase, the scale height is only as high as $\sim200-500\,\mathrm{pc}$ due to the lack of stellar feedback. In contrast, the 
associated break scale is around a kpc. For the \bk{MOL} gas, the disc scale height increases from the size of 
a grid cell at early times ($\sim$20pc; at this time there is essentially no gas \bk{in the MOL phase}) up to around 
$300\,\mathrm{pc}$ at late times. This is interesting on its own, as it shows that 
galactic dynamics can lift the gas towards such heights. However, the break scale is 
constantly around $100\,\mathrm{pc}$ and, thus, only matches the scale height at a short 
period in time. \\
\bk{We further derive the disc scale radius/length by fitting an exponential to the 
radial column density profiles. In case of the WNM, the scale radius is of the 
order of the disc extent, $\sim10\,\mathrm{kpc}$. This is due to the almost flat radial 
WNM gas distribution, where small-scale fluctuations are essentially averaged out (see also 
Fig.~\ref{fig:maps}). The MOL gas instead initially shows a scale radius of about a kpc, 
which reflects the inner core. With time, more and more clouds form throughout the disc 
and the scale radius increases due to non-negligible contribution at the disc edges. As 
the formed clouds merge and travel inwards over time, the scale radius shrinks again, but 
is still an order of magnitude larger than the corresponding break scale derived from the 
SPS.\\}
In order to characterize the typical separation of the \bk{MOL} gas, we follow \citet{Koch2019} and determine the two-point correlation function (TPCF) for the 
clouds in the image plane \citep[see also][]{Peebles}. The orange dash-dotted line highlights the 
lag distance, where the TPCF shows its largest value, which indicates the strongest correlation. This indicates the spacing between objects identified both in the WNM and \bk{denser MOL phase,} which decreases with time as more structures emerge. The TPCF lag shows some similarity with the break scale, but the agreement gets worse with time for the WNM phase and only trends with the shape for the \bk{MOL}.\\
Lastly, there is another scale within the image: the (projected) size of \Hone or H$_2$ gas clouds. We identify \bk{spatially} connected regions (clouds) in 3D space \bk{by using a 
simple clump finding algorithm, which checks whether adjacent grid cells have similar 
densities. If so, these cells are flagged as belonging to the same structure. We start 
with the maximum density within the simulation domain and walk through the 
\textit{density grid} down to the user-defined threshold density. We reject objects that 
are composed of less than ten grid cells. Our lower threshold density is 
\mbox{$n=100\,\mathrm{cm}^{-3}$} for the MOL phase and \mbox{$n=1\,\mathrm{cm}^{-3}$} 
for the WNM gas. For the latter phase, we further restrict the object temperatures to 
our specified temperature range. Finally, we} determine the size \bk{of the identified 
clumps} in two ways. First, by using the total volume of the identified objects ($V_\mathrm{tot}^{1/3}$), and second by determining the maximum distance from the center of mass of the objects. We then determine the distribution of sizes and extract the size at which this distribution peaks. \bk{The typical size of structures within both phases is 
quite similar, although the WNM structures appear smaller. This is due to the fact, that 
the WNM gas is more patchy, despite that smooth appearance in the column density maps. 
However, the spread in the size distribution is larger for the WNM structures with its 
maximum ranging around two kpc.}
Most interestingly, it is the typical size of the objects that matches best the break scale of the 2D input images. \bk{This indicates the dominance of dense, point-source like 
objects for the SPS.} The second striking feature is the lack of correlation between any scale and the break scale in the warm \Hone phase. Hence, the overall shape of the power spectrum and existence of a break is phase dependent and does not necessarily trace the same physics.\\
If there is no characteristic scale, then 
what does a break in the SPS of the WNM tell us? In Fig.~\ref{fig:spectra} we show kinetic energy 
power spectra for the WNM (left), CNM (middle) and \bk{MOL} phase (right). The three spectra per panel highlight the 
different components of the kinetic energy, i.e. by using the projected (and weighted) velocities. We 
further show the break scale and disc scale height of the example phases as solid and dashed vertical 
lines, respectively.\\
In general, all phases show the expected behaviour of a spatial regime, where the spectrum is almost flat, and 
a range where the power decreases with decreasing spatial scale. At late stages, 
however, the WNM phase shows a power-law shape over the entire range, which means that 
the typical energy injection scale is above the shown 20\,kpc. Strikingly, for the 
WNM and CNM phases, the transition from a plateau to a power-law regime in the energy spectrum occurs at the 
derived break scale calculated from the SPS. This is consistent with and further support for previous claims 
that the geometry of the turbulent cascade changes from \bk{disc-like} to 3D 
at the break scale. The isotropy of the turbulence below the break scale is nicely seen in the WNM phase\footnote{Note that we remove the bulk rotational velocity before the 
Fourier transform. Due to the small area filling fraction of the colder phases, the 
azimuthal spectra a slightly biased towards larger power.}. The disc-like turbulence is 
also highlighted by the fact that the kinetic energy of the vertical 
component reaches a plateau, while the azimuthal component increases further. In some 
cases, the radial component also saturates similarly to the vertical component. Further, it is clear from the panels that 
the disc scale height falls entirely within the isotropic regime and thus does not mark 
the outer scale of turbulence. Finally, as discussed above, the break scale does not 
match the transition from 2D to 3D turbulence in the \bk{MOL} phase as it is determined by 
the cloud size in this phase.

\subsection{\bk{Comparison with Previous Works}}
\bk{The SPS has been studied in numerical simulations before. \citet{Bournaud2010} argued 
that the break in their SPS was due to the disc thickness, based on a calculation of the 
Jeans length and the subsequent reason that Jeans length and scale height are usually of 
the same order of magnitude. As we have shown above, this argument does not hold and, 
instead, we here compare our SPS properties. The slopes of the time-averaged SPS on large and 
small scales by these authors were $\gamma_\mathrm{large}=-1.89$ and $\gamma_\mathrm{small}=-3.12$, respectively. It must be stated that the authors did not 
use a proper description for heating and cooling of the gas and thus only reached densities
of about $n\sim10^3\,\mathrm{cm}^{-3}$ (see their Fig.~1). The corresponding temperatures 
were a few tens of Kelvin so that our CNM phase matches best their data. As is seen from 
our Fig.~\ref{fig:slopes}, the CNM slopes are somewhat comparable on large scales, while the 
small scale slope is steeper towards later times. These small but recognizable differences might arise from different physics included. 
For instance, we here use a proper description for heating and cooling of the gas and, in 
addition, incorporate a strong magnetic field, whereas the simulations by \citet{Bournaud2010} were hydrodynamic ones.\\
More recently, \citet{Grisdale2017} studied the SPS for different types of galaxies and with 
or without stellar feedback from massive stars. They showed that stellar feedback is 
necessary to reproduce observed SPS from the THINGS survey. They presented evidence for 
feedback changing the location of a break in the SPS, but did not discuss this in detail. 
As can be seen, however, from their results, the change seems to depend on galaxy type. 
This, instead, might rather be due to the dynamics in the respective time period. \\
Our measured CNM slopes are best comparable with their feedback results, but significantly 
differ from their results without stellar feedback. The slopes derived for our MOL phase, 
which certainly resembles most the total column density (as in their simulations), are 
of similar value for the large scales, but much steeper on small scales. However, as 
identified in their Fig.~5, the SPS of galaxies without any form of feedback steepens by 
a large factor at decreased resolution. The slope on small scales now appears to be similar to 
their feedback simulations at higher resolution. Since their 18.3\,pc resolution is 
comparable to ours, this lends support to our findings  (see also our Fig.~\ref{fig:simulated_obs_breakscale} for resolution effects) and points towards 
effects of the resolution and filling fraction of the densest phase.
}

\subsection{Implications for Accretion-driven Turbulence}
\begin{figure*}
    \begin{tabular}{cc}
    \Large{L$=$Scale height}    &\Large{L$=$Break scale}\\
    \includegraphics[width=0.48\textwidth]{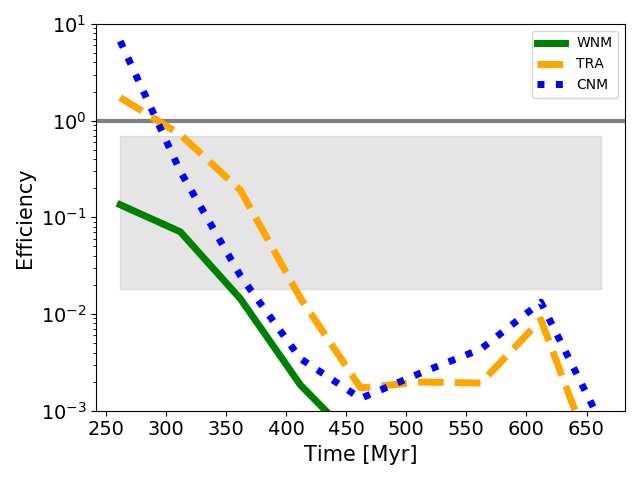} &\includegraphics[width=0.48\textwidth]{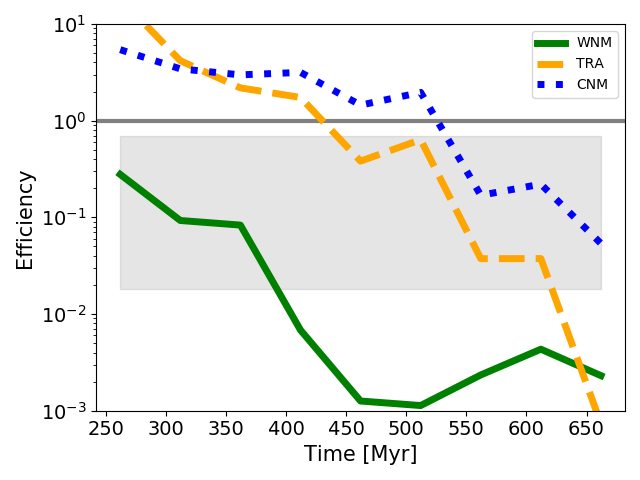}
    \end{tabular}
    \caption{Time evolution of the fraction of kinetic energy in the accretion 
    flow onto the galactic disc, which is necessary to 
    balance the natural decay of turbulent kinetic energy 
    in the ISM. The left panel shows the results, when $L$ from Eq. \ref{eq:energy_turb} is the
    scale-height, while the right panel uses the 
    break scale as estimated from the SPS. The 
    grey band presents estimates for various galaxies by 
    \citet{Klessen10}. An efficiency value $> 1$ indicates that turbulence cannot be sustained solely by accretion on to the galaxy. Hence, it is clear that 
    more energy has to be extracted from the accretion flow 
    to sustain turbulence in the individual phases. }
    \label{fig:accr_eff}
\end{figure*}

\bk{Above} we showed that, for all phases except the coldest phase \bk{(MOL)}, the 
break scale traces the transition from disc-like to 3D isotropic turbulence. Since the 
break scale is significantly larger than the disc scale height, this implies that the 
turbulence has already become isotropic well above the latter. We thus briefly discuss the 
subsequent implications for accretion-driven turbulence in galaxy discs. \bk{We emphasize 
here that the scale height, as measured from 
an edge-on projection, and the break scale, 
measured from a face-on map, provide the two 
limiting cases for the size scales from 
different galaxy inclinations. A 
note on inclination effects on the SPS 
is provided in the appendix below.}\\
\citet{Klessen10} define the fraction of accretion energy that is required to 
sustain the turbulent cascade within the ISM as 
\beq\label{eq:energy_efficiency}
\epsilon = \left|\frac{\dot{E}_\mathrm{turb}}{\dot{E}_\mathrm{accr}}\right|,
\eeq

where 
\beq\label{eq:energy_turb}
\dot{E}_\mathrm{turb}\sim-\frac{1}{2}\frac{M\sigma_\mathrm{turb}^3}{L}
\eeq
is the decay rate of turbulent energy, with $M$ being the total mass (in the respective phase), $\sigma_\mathrm{turb}$ the turbulent velocity dispersion and $L$ defining a 
characteristic scale of the system. On the other hand, the accretion energy is 
given by 
\beq\label{eq:energy_accr}
\dot{E}_\mathrm{accr}\sim\frac{1}{2}\dot{M}_\mathrm{accr}v_\mathrm{accr}^2.
\eeq
Here, $\dot{M}_\mathrm{accr}$ is the mass accretion rate and $v_\mathrm{accr}$ the 
corresponding accretion velocity. Hence, the smaller $\epsilon$, the smaller the 
fraction of required accretion energy needed to sustain the turbulent motions in the 
galaxy disc.\\
To determine the efficiency, we calculate the disc scale height and estimate the 
mass and velocity dispersion in a volume limited by the scale height ($M$ and $\sigma_\mathrm{turb}$ respectively). At the 
same time, we measure the mass accretion rate and accretion velocity ($\dot{M}_\mathrm{accr}, v_\mathrm{accr}$) in a cylindrical volume with height of one kpc, width of 
$D=2R=20\,\mathrm{kpc}$, centered on the center of the 
galaxy,
and having its lower boundary at the scale height. The accretion rate is simply 
given as $\int{\varrho v_\mathrm{cell,a}dA}$, where $v_\mathrm{cell,a}$ is the 
velocity of the cell directed towards the disc midplane. These 
calculations are performed for gas in different phases.\\
In the \ita{left} panel of Fig.~\ref{fig:accr_eff} we show the fraction of required accretion energy 
(here called \ita{efficiency}) for three phases as a function of time. The grey 
shaded area presents the range of efficiencies for various galaxies from the 
THINGS survey \citep{Walter08} as estimated 
by \citet{Klessen10}. An efficiency value $>1$ indicates that accretion onto the 
galaxy is insufficient to sustain turbulent motions within the ISM. 
Initially, the turbulence in the colder phases cannot be driven by 
accretion onto the galaxy, simply because the disc has not fragmented yet and the amount of material in these phases is small. In 
contrast, the warm phase shows an efficiency of 10\,\%, which 
indicates that accretion of matter in this phase is indeed able to 
sustain turbulent motions in the WNM. The values for the required 
fraction also fall in the limits estimated for the THINGS survey.\\
Over time, the needed fraction of accretion energy quickly drops off 
to tiny values. Since we do not include feedback in this study, 
accretion flows seem to easily sustain turbulence in the galaxy\footnote{Whether this is the dominant source of turbulence in our simulated galaxies remains to be studied.}. The inclusion of 
feedback, which opposes accretion flows, should raise the efficiency 
values more towards observed values. Note further that the 
required fractions do not differ too much between the various phases.\\
In the \ita{right} panel of Fig.~\ref{fig:accr_eff} we show the same 
quantity, but now we use the break scale as characteristic scale. 
The results change dramatically by showing a different time evolution, as well as a stronger phase dependency. The efficiency of the 
WNM phase changes only mildly, indicating that the turbulence in the 
warm phase of the galactic ISM can easily be sustained by accretion 
flows. However, the colder phases now show values $>1$ for almost 
two entire orbits at $R=8\,\mathrm{kpc}$. This highlights that 
accretion of cold material is not sufficient to sustain the 
turbulent motions in the same phase within the galaxy. Here, 
additional mechanisms are needed. At late stages, the efficiency 
finally drops to below unity, but the rate of decrease is larger 
for the transitional phase such that it ends up having smaller 
efficiencies than the WNM phase. The cold \Hone phase ends up at 
required fractions of about 10\,\%.\\
To summarize, the break scale serving as the outer scale of 
isotropic turbulence in galaxy discs implies that accretion flows 
onto the galaxy are not as efficient in driving turbulence in the 
galactic ISM and show a stronger phase dependence. We, however, mention the caveat of not taking into 
account the energy input from the warmer phases, i.e. of the WNM when determining the efficiency of the CNM.

\subsection{Implications for the SPS in Observations}

A recent study by \cite{koch2020} suggests the SPS break scale, if present, to be in many cases a tracer of the PSF response of the observing instrument rather than a reflection of the disc scale height. In order to test the influence of instrumental systematics, such as the PSF response and noise, on our ability to detect a genuine break scale, we seek to recreate observational effects by simulating an observation of the face-on galaxy studied herein.\\
\np{We select the evolved 612 Myr time step of the NEU phase to assess the effects of instrumental noise and finite PSF response on the shape of the SPS. Normally distributed noise at various signal-to-noise ratios (SNR) is added to the simulated column density image such that the standard deviation is proportional to the peak intensity. We then down-sampled the noisy image by factors of $2^n$ for $n=0,1,2,3,4$ and convolved with a Gaussian kernel of 5 pixels.} Down-sampling to progressively coarser grids ensures the resulting pixel sizes in our final maps do not oversample the expected angular scales in the convolved images, while the 5 pixel FWHM of the Gaussian kernel ensures an adequate number of resolution elements across the beam. The angular resolution of these down-sampled maps are determined by placing the simulated galaxy at a distance of 1 Mpc. As we have done with the original images, we process the resultant images with {\tt Turbustat} to calculate the break scales. We exclude results that fail to fit a broken power-law.

\begin{figure*}
    \centering
    \includegraphics[width=\textwidth]{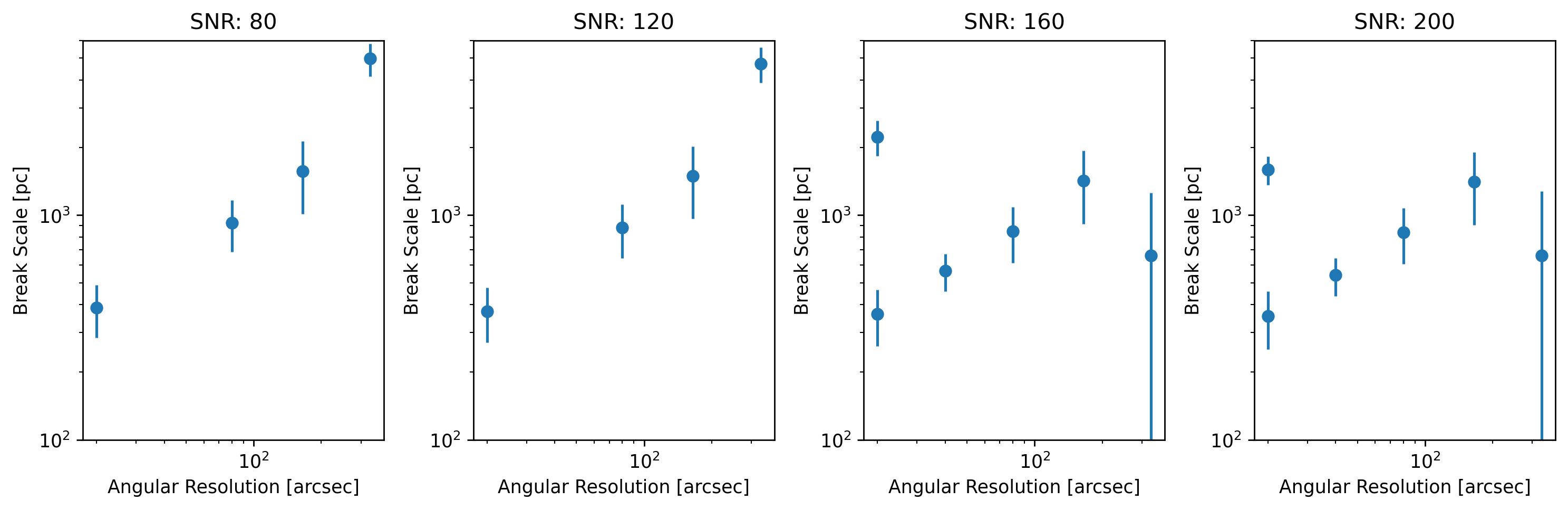} 
    \caption{The measured break scale of the \bk{NEU phase} assessed for differing beam sizes and additional instrumental noise. On the left hand axis is the break scale. The bottom axis of each column measures the angular resolution of the image. This was determined from how much the image was down-sampled prior to convolution with a gaussian kernel (e.g. 165" corresponds to an image down-sampled to 1/16th of the original size in both height and width). The top label on each column indicates the SNR of the simulated image. The errorbars are the measured uncertainty on the break scale calculation. Images with noise and angular resolution combinations that failed to be fit with a broken power law were excluded from the plot. The plots show a marked upwards trend in measured break scale as a function of beam size.}
    \label{fig:observations}
\end{figure*}

The results in Fig. \ref{fig:observations} show the resulting break scales as a function of angular resolution for \np{several SNRs.} We see a consistent upwards trend of the observed break scale with increasing angular scale/down-sampling. These results demonstrate as the smaller spatial scales are smoothed out through resampling, their power is reduced and the break scale measured shifts to larger scales; accordingly, as the down-sampling becomes higher, larger scales become smoothed out and the break scale shifts to even higher values. \np{The increase in break scales at low angular resolutions for the SNR 160 and SNR 200 correspond to the cases where no convolution is performed after the addition of the noise. These large increases in the \bkk{estimates} of the break scale indicate that small-scale fluctuations in the noise, such as noise peaks near $\sim$3$\sigma$, significantly influence the overall shape of the SPS by adding power towards the smaller scales.} We also note that we yielded similar results to those in Fig. \ref{fig:observations} from following a more straightforward method of convolving the images with Gaussians of varying pixel widths.

These results suggest there is a strong relationship between the minimum resolved spatial scales/beam size and the break scale observed. There is a trend towards large break scales that is comparable across noise values\np{,} indicating the angular resolution of the beam is the dominant driver of the observed break scales here. Consequently, we conclude that, while unavoidable, the addition of any noise to the observations will adversely affect the recovery of the inherent break scale of the ISM. Additionally the break scale recovered will be biased towards larger values \np{at lower angular resolutions.}



The trends in Fig.~\ref{fig:observations} indicate that observed break scales of external galaxies are significantly influenced by observational systematics. Following ~\citet{koch2020}, we attempt to recover the break scales measured in the original images by fitting a broken power-law model that accounts for the noise and forward models the PSF response. We adopt the smoothly broken power law model 
\begin{equation}\label{eq:broken_pl}
    P_{\rm broken}(k) = A \left(\frac{k}{k_b}\right)^{-\beta}\left\{\frac{1}{2}\left[1+\left(\frac{k}{k_b}\right)^{1/\Delta}\right]\right\}^{\beta-\beta_2}, 
\end{equation}

where $A$ is the amplitude of the power, $k_b$ is the characteristic break scale, and $\beta$ and $\beta_2$ are the spectral indices above and below the break, respectively. The $\Delta$ parameter controls the smoothness of the break. Through visual inspection, we find setting $\Delta$=0.15 accurately captures the smoothness of our measured SPS profiles. The PSF response is characterized in our fit by first multiplying $P_{\rm broken}$ by the SPS of our simulated Gaussian kernel placed on the same coarse grid. We account for the noise by introducing an additional term, $C$, that multiplies the power spectrum of an image of normally distributed noise, $P_{\rm noise}(k)$, according to the specific SNR in our simulated image:
\begin{equation}\label{eq:broken_pl_psf}
P(k) = P_{\rm PSF}\cdot P_{\rm broken}+C\cdot P_{\rm noise}(k).
\end{equation}

We utilize the {\tt PYMC3} python package to run Markov chain Monte Carlo (MCMC) sampling of the parameter space of our model to ensure our uncertainty in the fitted break scale is fully characterized and robust. Given the wide range of possible values for the amplitudes $A$ and $C$, we sample in $log_{10}$ space and adopt uniform priors:

\begin{equation}\label{eq:A_prior}
log_{10}A \sim \mathcal{U}(-20, 20)
\end{equation}
\begin{equation}\label{eq:C_prior}
\gamma \sim \mathcal{U}(-20, 20)
\end{equation}
\begin{equation}\label{eq:beta_prior}
\beta \sim \mathcal{U}(0, 10)    
\end{equation}
\begin{equation}\label{eq:beta_2_prior}
\beta_2 \sim \mathcal{U}(0, 10)    
\end{equation}
\begin{equation}\label{eq:k_b_prior}
k_b \sim \mathcal{U}(k_{\rm min}, k_{\rm max}), 
\end{equation}

where $k_{\rm min}$ and $k_{\rm max}$ represent the spatial frequencies associated with 1/2 of the input image size and size of the beam at a given angular resolution, respectively. Finally, we assume the points in the SPS profiles to be independent samples of a normal distribution with width equal to the standard deviation of power values in each bin. These samples are again drawn in $log_{10}$ space to account for potentially large variations.

\begin{figure}
    \centering
    \includegraphics[width=0.5\textwidth]{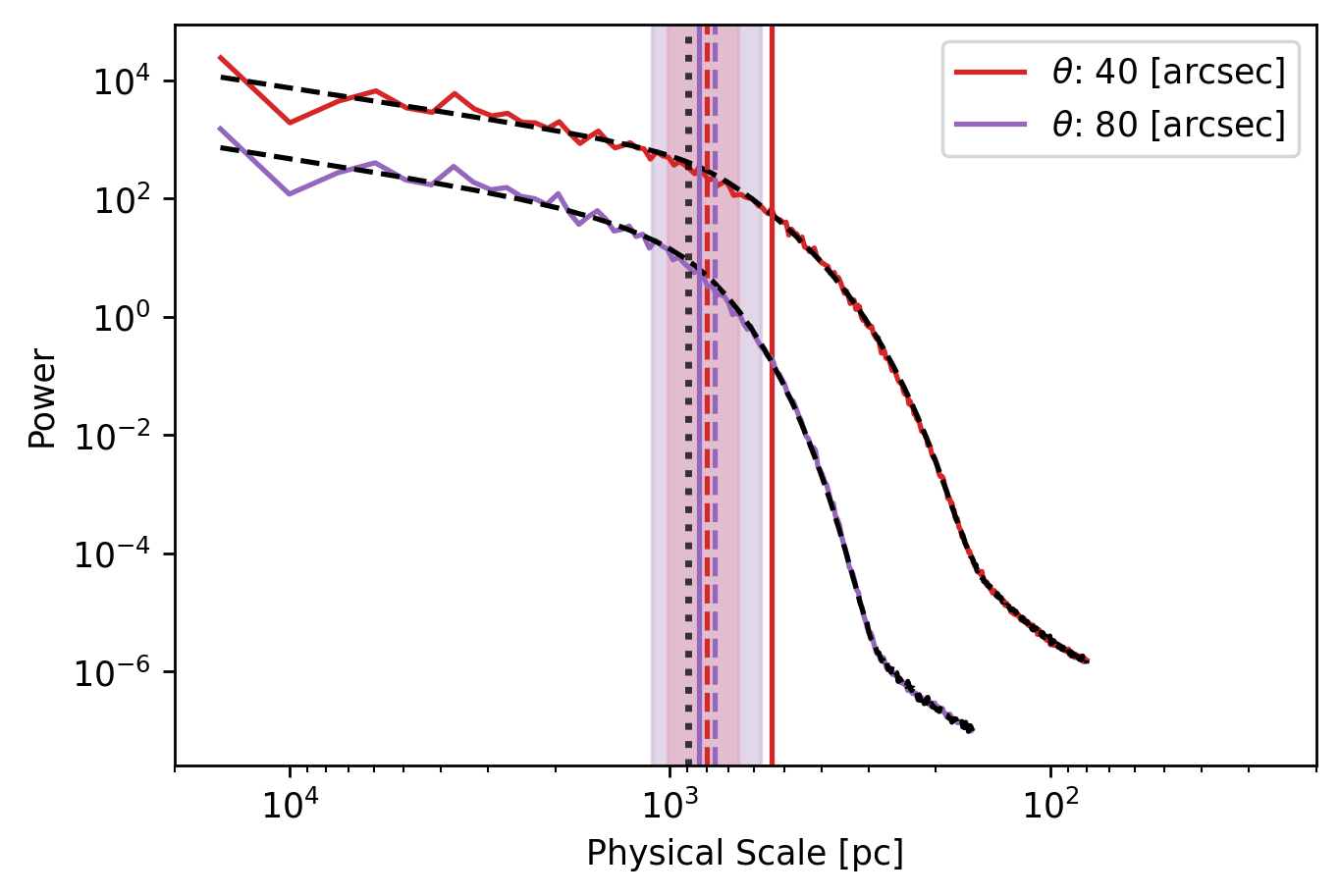}
    \caption{Power spectra measured for the simulated observations of the \np{NEU} phase with SNR 200 at \np{intermediate} angular resolutions. The solid vertical lines denote the break scale calculated by {\tt Turbustat} when the noise and beam are not accounted for. The dashed vertical lines and shaded region denote the break scale estimate and standard deviation as determined by the fit to a smoothly-broken power law model that forward models the PSF and accounts for instrumental noise, which are overlaid on each profile as dashed black lines. The vertical black dotted line shows the original break scale for this phase at a timestep of 612 Myr. The discrepancy between the measured break scales demonstrates the need to account for observational systematics, such as the PSF, when extending this analysis to real observations.}
    \label{fig:simulated_obs_breakscale}
\end{figure}
Figure~\ref{fig:simulated_obs_breakscale} shows the fit of Equation~\ref{eq:broken_pl_psf} to the SPS profiles of the \np{NEU} phase with a measured break scale from the {\tt Turbustat} at a \np{SNR} of 200. The fitted break scales, when accounting for the contribution of noise and forward modeling the PSF response, are much closer to the true break scale measured from the original simulated images for \np{these two intermediate angular resolutions. The profiles at other angular resolutions are not shown for clarity but we note they are similarly well fit. However, the uncertainty in the fitted break scale increases with decreasing angular resolution, demonstrating that the true break scale becomes difficult to recover past intermediate angular resolutions. These results demonstrate that it is vital to account for observational systematics in power spectra analysis of real observations. The effects of the observing systematics can approximately be extended to the dust distribution, given the excellent spatial coincidence between \Hone and FIR dust emission observed in the Milky Way and nearby galaxies \citep{compiegne2011, Clark2019, stanimirovic2000}}.

\section{Summary}\label{sec:summary}
In this study we presented results from global galaxy simulations. The specific 
target was to analyze the time and phase dependency of the spatial power 
spectrum. To achieve this task, we split the total gas into regimes of 
different temperature and studied the time evolution of the break scale in the 
spatial power spectrum as well as of its slopes on large and small scales. Our 
key findings can be summarized as follows:
\begin{itemize}
    \item[i)]   The break scale in the SPS is phase and mildly time dependent, but the combined 
    spectrum is dominated by the coldest, densest phase.
    \item[ii)]  The large-scale slope (i.e. the slope on scales larger than the 
    break scale) does not vary too much between various phases.
    \item[iii)] The small-scale slope shows a slight phase dependence and is 
    broadly consistent with compressive turbulent motions.
    \item[iv)]  The break scale of the SPS of face-on galaxies does not trace the 
    disc scale height and thus does not provide a hint towards the extent along 
    the line of sight. Instead, the physics buried within the break scale 
    depends on phase. For the warmer phases, the break scale rather traces the 
    transition from 2D disc-like to 3D isotropic turbulence. In the cold phase, it 
    traces the typical size of molecular clouds.
    \item[v)]  As the break scale traces a transition in turbulence, it affects 
    the conversion of accretion energy to random turbulent kinetic energy in the 
    ISM of galaxies. As such, the required energy fraction to sustain turbulent 
    motions in late-type galaxies is increased.
    \item[vi)] The accurate measurement of an underlying break scale in an actual external galaxy is highly dependent on instrumental effects, such as the finite PSF and noise. Consistent with the results of \citet{koch2020}, we demonstrate that these must be accounted for in any power spectra analysis of real observations.
\end{itemize}
Our numerical investigation provides further arguments against the break scale tracing the 
third dimension of face-on galaxies. In addition, while a major limitation of our simulations is 
the lack of stellar feedback from e.g. supernovae, it is unlikely to strongly alter our conclusion 
regarding the cloud size. \citet{Seifried2018} show that external supernovae do not drive turbulence 
within molecular clouds and \citet{Koertgen16} emphasize that this form of feedback is unlikely to 
disintegrate the clouds. Hence, typical cloud sizes from our current study should persist even in 
runs with active stellar feedback. Stellar feedback will thus primarily affect the mass fractions of 
gas in the various phases with temperatures $T>50\,\mathrm{K}$ and subsequently the power spectra 
of these. Specifically, we expect only the break scale to shift towards larger scales, but the 
physics traced by it remain the same.

\section*{Acknowledgement}
\bk{We thank the anonymous referee for a timely and constructive report.}
We acknowledge Paris-Saclay University's Institut Pascal program "The Self-Organized Star Formation Process" and the Interstellar Institute for hosting discussions that nourished the development of the ideas behind this work. BK acknowledges discussions with 
R.~Banerjee and P.~Trivedi. BK thanks for funding from the DFG grant
BA 3706/15-1 and via the Australia-Germany Joint
Research Cooperation Scheme (UA-DAAD). NP acknowledges that this research was supported by the Australian Research Council (ARC) through grant DP190101571. NK-S is supported by an Australian Government Research Training Program (RTP) Scholarship. The \textsc{flash} code was in part developed by the DOE-supported ASC/Alliance Center for Astrophysical Thermonuclear Flashes at the University of Chicago. The work was supported by the North-German Supercomputing Alliance (HLRN) under project ID hhp00050.

\section*{Data availability}
The data underlying this article will be shared on reasonable
request to the corresponding author.

\begin{appendix}
\section{\bk{Dependencies of the break scale}}
\bk{Apart from characteristic physical scales of the galaxies, the break scale further shows
dependence on the PSF of the instrument. Hence, the break scale might also depend on 
other quantities of the system or the image. In Fig.~\ref{fig:app_a} we show the derived 
break scales as a function of mass fraction and of the image filling fraction. From both 
panels it is clear that there is no clear correlation between these two quantities. This 
reveals that the break scale traces physics, hidden in the image, but no systematics of 
the image itself.
\begin{figure*}
    \centering
        \includegraphics[width=0.5\textwidth]{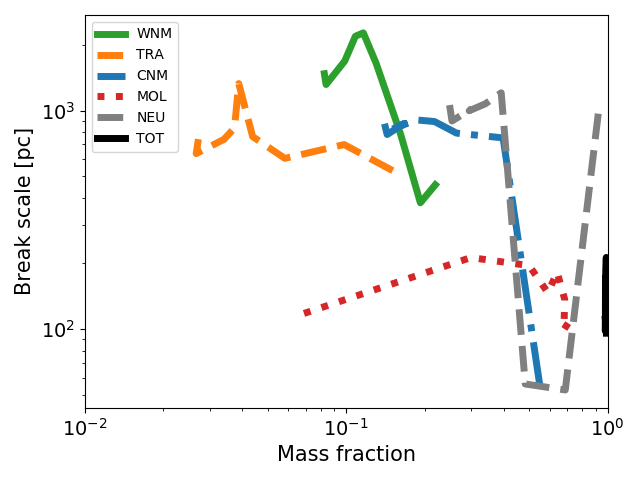}\includegraphics[width=0.5\textwidth]{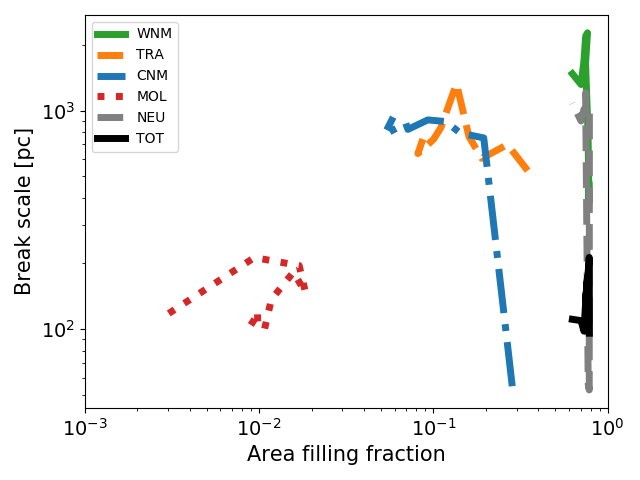}
        \caption{\bk{Dependence of the break scale on the mass fraction (left) and the 
        image filling fraction (right). It is seen that there is no clear correlation 
        between the quantities, so that an effect by these quantities can be ruled out.}}
        \label{fig:app_a}
\end{figure*}
}
\section{\bk{Inclination effects}}
\bk{A bias is certainly introduced via the inclination of the observed galaxy. \citet{Grisdale2017} studied two inclinations, namely face-on (i.e. $i=0^{\circ}$) and 
$i=40^{\circ}$, where the latter was the average disc inclination of their observational 
control data from the THINGS survey. Here, we use the two limiting inclinations of $i=0^{\circ}$ and $i=90^{\circ}$, corresponding to face-on and edge-on discs. In Fig.~\ref{fig:app_b} we show the SPS for various time steps. The spectra are normalized to 
the total spectral power for better comparison. In the un-normalized case, the edge-on 
SPS contains significantly more power on all scales, since we are integrating through 
the entire disc, encompassing almost 20\,kpc of 'disc material'. The vertical lines 
denote the individual break scales and the derived scale height. The scale height 
increases with time and does not necessarily match any of the measured break scales. 
While the break scales trace typical scales of the column density map, the scale height 
is much more sensitive to gas, which has been lifted to large heights from galactic 
dynamics. \\
Apart from the large difference in the break scale for these two limiting cases of disc 
inclination, we see a surprisingly large similarity of the slopes above and below the 
individual break scales. Hence, deriving turbulent dynamics from the SPS seems to not 
strongly depend on disc inclination. The 
derived efficiency shown in Fig.~\ref{fig:accr_eff} are thus limiting 
cases and inclinations $0^{\circ}<i<90^{\circ}$ will thus lie between these two cases in good 
agreement with the findings by \citet{Klessen10}.
\begin{figure*}
    \centering
        \includegraphics[width=\textwidth]{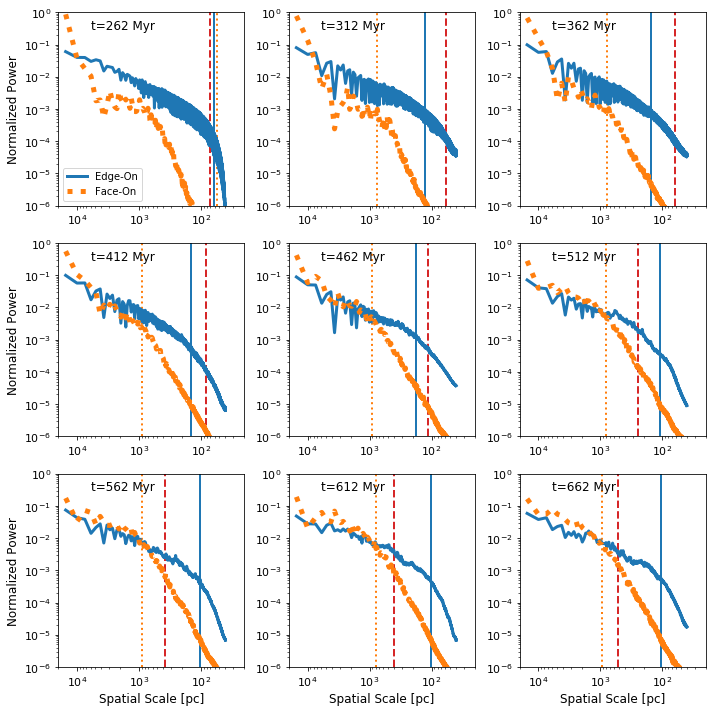}
        \caption{\bk{Normalized SPS for the edge-on and face-on view at various times. 
        The vertical lines represent the derived break scales as well as the scale height (red dashed). These two views represent the two extreme cases for the disc 
        inclination. For the un-normalized SPS, the edge-on SPS dominates in power due to 
        integration through the entire radial extent of the galaxy. Note the close 
        similarity of the slopes above and below the break scale.}}
        \label{fig:app_b}
\end{figure*}
}
\end{appendix}

\bibliography{astro_mod, additional_refs} 
\bibliographystyle{mn2e}
\end{document}